\documentclass[letterpaper,journal]{IEEEtran}
\usepackage{amsmath,amsfonts}
\usepackage{algorithmic}
\usepackage{algorithm}
\usepackage{array}
\usepackage{textcomp}
\usepackage{stfloats}
\usepackage{url}
\usepackage{verbatim}
\usepackage{hyperref}
\usepackage{graphicx}
\usepackage[caption=false,font=normalsize,labelfont=sf,textfont=sf]{subfig}
\usepackage{cite}
\usepackage{amsmath}
\usepackage{subcaption}
\usepackage{booktabs,multirow,makecell}
\usepackage{siunitx}
\usepackage{caption}
\usepackage[table]{xcolor} % 可选，用于行底色

\begin{document}

\title{Multi-scale Cascaded Foundation Model for Whole-body Organs-at-risk Segmentation}

\author{Rui Hao, Dayu Tan, Qiankun Li, Chunhou Zheng, Weimin Zhong and Zhigang Zeng, \IEEEmembership{Fellow, IEEE}

\thanks{The work was supported by the National Key R\&D Program of China under Grant 2021ZD0201300, the Foundation for Outstanding Research Groups of Hubei Province of China under Grant 2025AFA012, the 111 Project on Computational Intelligence and Intelligent Control under Grant B18024, and the Postdoctoral Fellowship Program of CPSF under Grant Numbers GZB20250428 and 2025M771702.(\emph{Corresponding author: Dayu Tan and Zhigang Zeng.})
}
\thanks{Rui Hao and Zhigang Zeng are with the School of Artificial Intelligence and Automation, Huazhong University of Science and Technology, Wuhan 430074, China, also with the Institute of Artificial Intelligence, Huazhong University of Science and Technology, Wuhan 430074, China, also with the Hubei Key Laboratory of Brain-Inspired Intelligent Systems, Huazhong University of Science and Technology, Wuhan 430074, China, and also with the Key Laboratory of Image Processing and Intelligent Control (Huazhong University of Science and Technology), Ministry of Education, Wuhan 430074, China (e-mail: zgzeng@hust.edu.cn).
}
\thanks{Dayu Tan and Chunhou Zheng are with the Key Laboratory of Intelligent Computing and Signal Processing, Ministry of Education, Anhui University, Hefei 230601, China.}
\thanks{Qiankun Li is with the College of Computing and Data Science (CCDS), Nanyang Technological University, 639798, Singapore.}
\thanks{Weimin Zhong is with the East China University of Science and Technology, Shanghai 200237, China.}
}

\markboth{}%
{Hao \MakeLowercase{\textit{et al.}}: Multi-scale Cascaded Foundation Model for Whole-body Organs-at-risk Segmentation}


\maketitle

\begin{abstract}
Accurate segmentation of organs-at-risk (OARs) is vital for safe and precise radiotherapy and surgery. Most existing studies segment only a limited set of organs or regions, lacking a systematic treatment of OARs segmentation. We present a Multi-scale Cascaded Fusion Network (MCFNet) that aggregates features across multiple scales and resolutions. MCFNet consists of a Sharp Extraction Backbone for the downsampling path and a Flexible Connection Backbone for skip-connection fusion, strengthening representation learning in both stages. This design improves boundary localization and preserves fine structures while maintaining computational efficiency, enabling reliable performance even on low-resolution inputs. Experiments on an NVIDIA A6000 GPU using 36,131 image–mask pairs from 671 patients across 10 datasets show consistent robustness and strong cross-dataset generalization. An adaptive loss-aggregation strategy further stabilizes optimization and yields additional gains in accuracy and training efficiency. Through extensive validation, MCFNet outperforms existing methods, excelling in organ segmentation and providing reliable image-guided support for computer-aided diagnosis. Our solution aims to improve the precision and safety of radiotherapy and surgery while supporting personalized treatment, advancing modern medical technology. The code has been made available on GitHub: \href{https://github.com/Henry991115/MCFNet}{\textcolor{blue}{https://github.com/Henry991115/MCFNet}}.
\end{abstract}


\section{Introduction}
\IEEEPARstart{M}{edical} image segmentation aims to accurately separate different tissues, organs, or lesion areas, supporting clinical tasks like disease analysis, lesion detection, and treatment planning \cite{ref1,ref2}. Accurate segmentation of malignant tumors is crucial for early cancer diagnosis, treatment planning, and efficacy evaluation. However, tumors often have complex shapes, blurred boundaries, and are closely connected to surrounding tissues, posing challenges for segmentation \cite{ref3,ref4,ref5,ref6}. Identifying and locating OARs is vital in radiation therapy, helping minimize damage to healthy tissues and improving treatment outcomes \cite{ref7,ref8,ref9}. Tumor and OAR segmentation are essential for precise treatment planning, enabling targeted radiation while protecting healthy tissues \cite{ref10,ref11}.

\begin{figure}[!t]
\centerline{\includegraphics[width=1.0\linewidth]{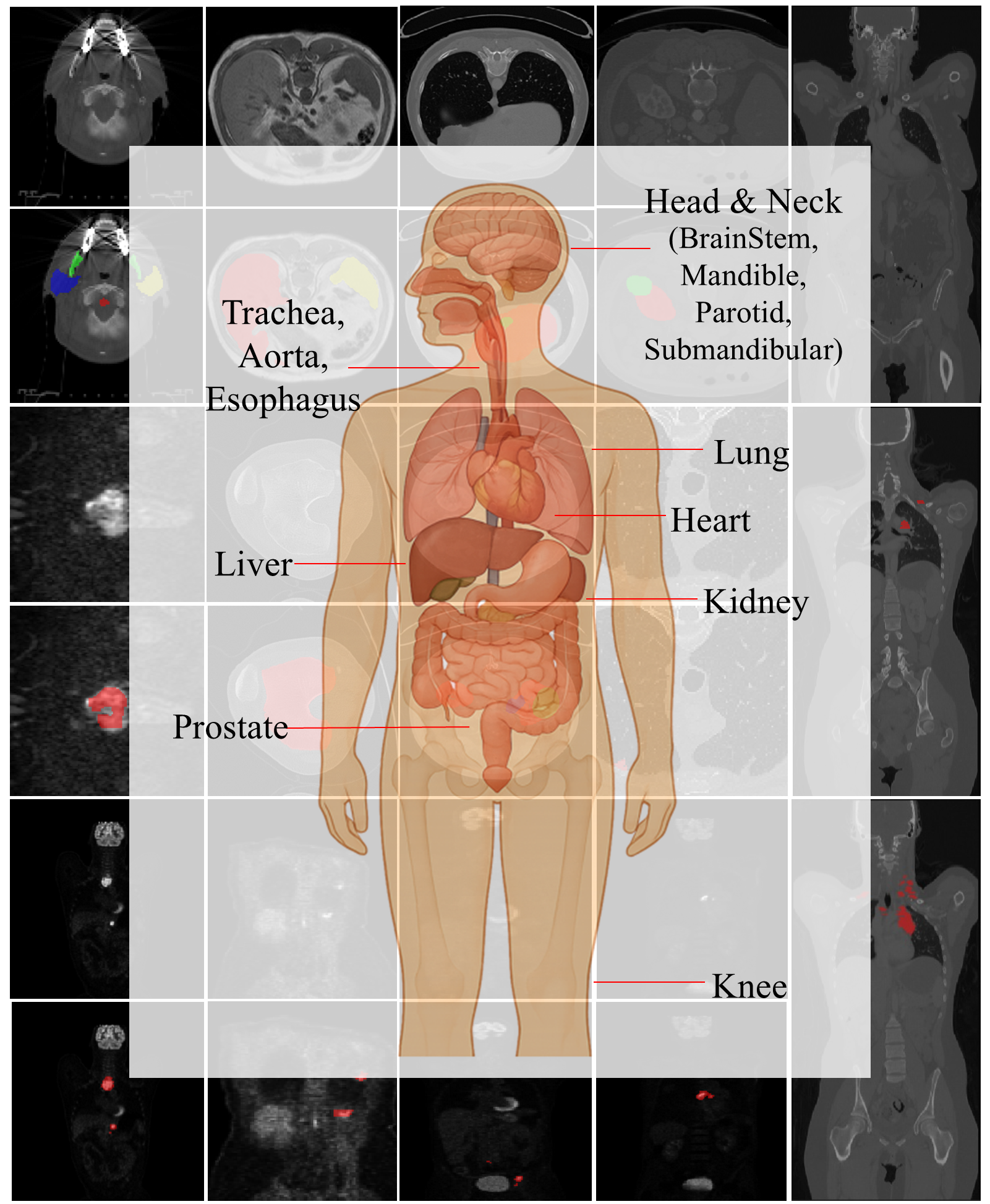}}
\caption{MCFNet is trained on ten diverse datasets to perform whole-body organs-at-risk (OARs) segmentation, covering the head \& neck, thorax, abdomen, prostate, and femur regions.}
\label{wholeBody}
\end{figure}
With the rise of deep learning, traditional methods constrained by human and financial costs are being replaced \cite{ref12,ref13,ref14,ref15}. Deep learning models can automatically extract complex features from images, enabling efficient OARs contour identification and further medical analysis \cite{ref16,ref17,ref18}. However, most current segmentation models require substantial GPU resources for training and large-scale annotated datasets, which are time-consuming and resource-intensive to create \cite{ref19,ref20}.

Most current deep learning segmentation models are based on UNet \cite{ref21} and Transformer \cite{ref22} architectures. UNet is known for its simple structure that effectively captures local features, while Transformers excel at capturing global information \cite{ref23,ref24,ref25}. Combining UNet's local feature extraction with the Transformer's global information capture improves segmentation accuracy and robustness by considering both local details and the overall structure in images. This combination enhances medical image analysis, supporting progress in the field \cite{ref26,ref27,ref28}. However, the segmentation performance of existing methods is often limited by the application scenario, with weak robustness and generalization abilities. High-performance models such as TransUNet \cite{ref29}, Swin-Unet \cite{ref30}, and MISSFormer \cite{ref31} perform well in segmenting organs in the heart and abdominal regions, while UCTransNet \cite{ref32} can segment glands and cell nuclei. Nevertheless, in our actual experiments, we find that the segmentation performance of these methods varies significantly when applied to datasets with different image characteristics \cite{ref33}.

In this paper, we propose MCFNet, a multi-scale cascaded fusion framework for robust segmentation of OARs and lesions under diverse imaging protocols and anatomical regions. MCFNet synchronizes multi-scale and multi-resolution inputs through two complementary branches: the Flexible Connection Backbone (FCB) for coarse global semantics and the Sharp Extraction Backbone (SEB) for fine boundary details. A Linear Attention Transformer is embedded in FCB to model long-range dependencies efficiently, while SEB focuses on high-frequency structures. An adaptive multi-scale loss aggregation strategy further enhances accuracy and training stability.  As shown in Fig. \ref{wholeBody}, we evaluate MCFNet on ten datasets spanning head-neck, thorax, abdomen, prostate, and femur regions. Extensive experiments and ablation studies demonstrate its strong robustness, generalization, and parameter efficiency.

The main contributions of our work are as follows:

\begin{itemize}
\item[$\bullet$] We propose MCFNet, a novel multi-scale cascaded network that integrates the SEB and FCB to encode and fuse features across scales and resolutions. Feature fusion is performed at four skip connections and the bottleneck layers, enabling both boundary fidelity and semantic consistency. Comprehensive evaluations on 10 heterogeneous datasets show that the proposed framework is robust, computationally efficient, and generalizes well.

\item[$\bullet$] We develop a Linear Attention Transformer Block (LAT) within the FCB to enhance feature extraction along the channel and spatial axes. LAT reduces quadratic complexity to linear, thereby accelerating convergence and improving scalability. This design enables the FCB to capture global dependencies with fewer parameters, achieving a favorable accuracy–efficiency trade-off.

\item[$\bullet$] We propose an Adaptive Multi-scale Feature-Mixing Loss Aggregation (Adaptive-MFA) strategy, which dynamically generates new prediction maps by combining multi-scale outputs. This adaptive aggregation stabilizes training and further improves segmentation performance.
\end{itemize}

The structure of this study is as follows: Section \ref{sec:related work} discusses related work. The methodology of our proposed model is detailed in Section \ref{sec:proposed method}. The experiment configuration is introduced in Section \ref{sec:experiments}. Section \ref{sec:experiments} describes the results and discussion, and the effectiveness of the proposed method is validated. Finally, Section \ref{sec:conclusion} presents the summarization of our research conclusions.

\begin{figure}[!t]
\centerline{\includegraphics[width=1.0\linewidth]{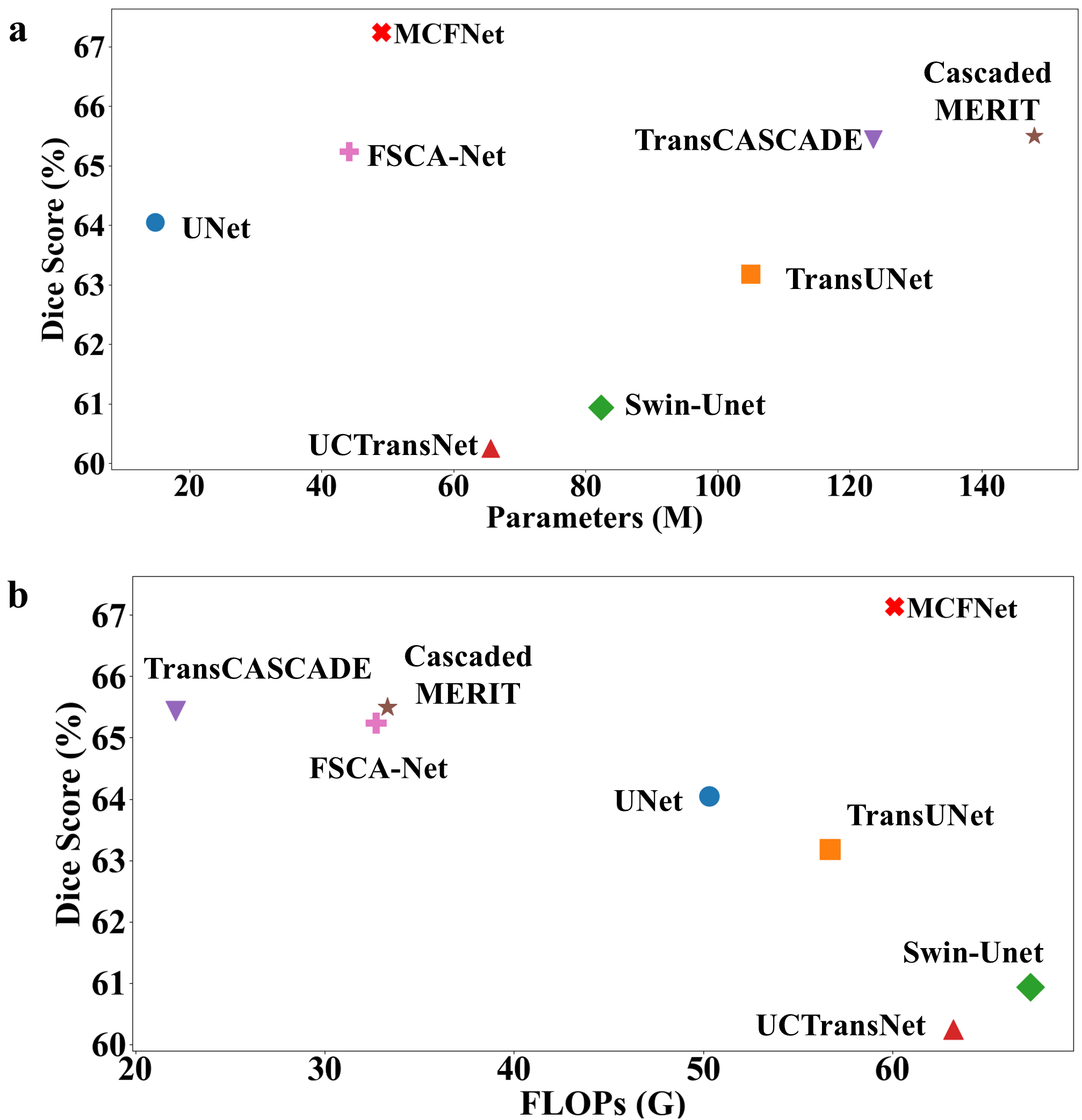}}
\caption{Discussion on Model Complexity on the CPCGEA Dataset. (a) Visualization of model performance and parameter size. (b) Visualization of model performance and FLOPs.}
\label{modelParamFLOPs}
\end{figure}

\section{Related Work}
\label{sec:related work}
This section introduces common architectures for medical image segmentation and cascade networks.

\subsection{Medical Image Segmentation}
Medical image segmentation, which involves accurately separating different structures like organs and lesions, is essential for medical image analysis and diagnosis. The introduction of encoder-decoder-based U-shaped networks, such as UNet, has significantly advanced this task, with subsequent improvements like UNet++ \cite{ref34} and UNet 3+ \cite{ref35} enhancing skip connection mechanisms for better feature aggregation.

In recent years, Transformers have gained attention in medical image segmentation. Although initially designed for NLP, Transformers excel in computer vision tasks due to their powerful modeling and flexibility \cite{ref36}. Unlike CNNs, Transformer models employ self-attention to model long-range dependencies on a global scale. ViT~\cite{ref37} introduce Transformers to image classification, using fixed-size patches to segment images, treating them as sequences for processing. This idea has been adapted to medical image segmentation, exemplified by the TransUNet model, which combines UNet's local feature extraction with the global context modeling of Transformers. The Swin Transformer, a hierarchical model with sliding windows, further boosts feature extraction and efficiency. The Cascaded MERIT model by Rahman et al.~\cite{ref38} has been shown to improve segmentation accuracy and stability using multiscale features and cascaded attention decoding.

\begin{figure*}[!t]
\centerline{\includegraphics[width=1.0\linewidth]{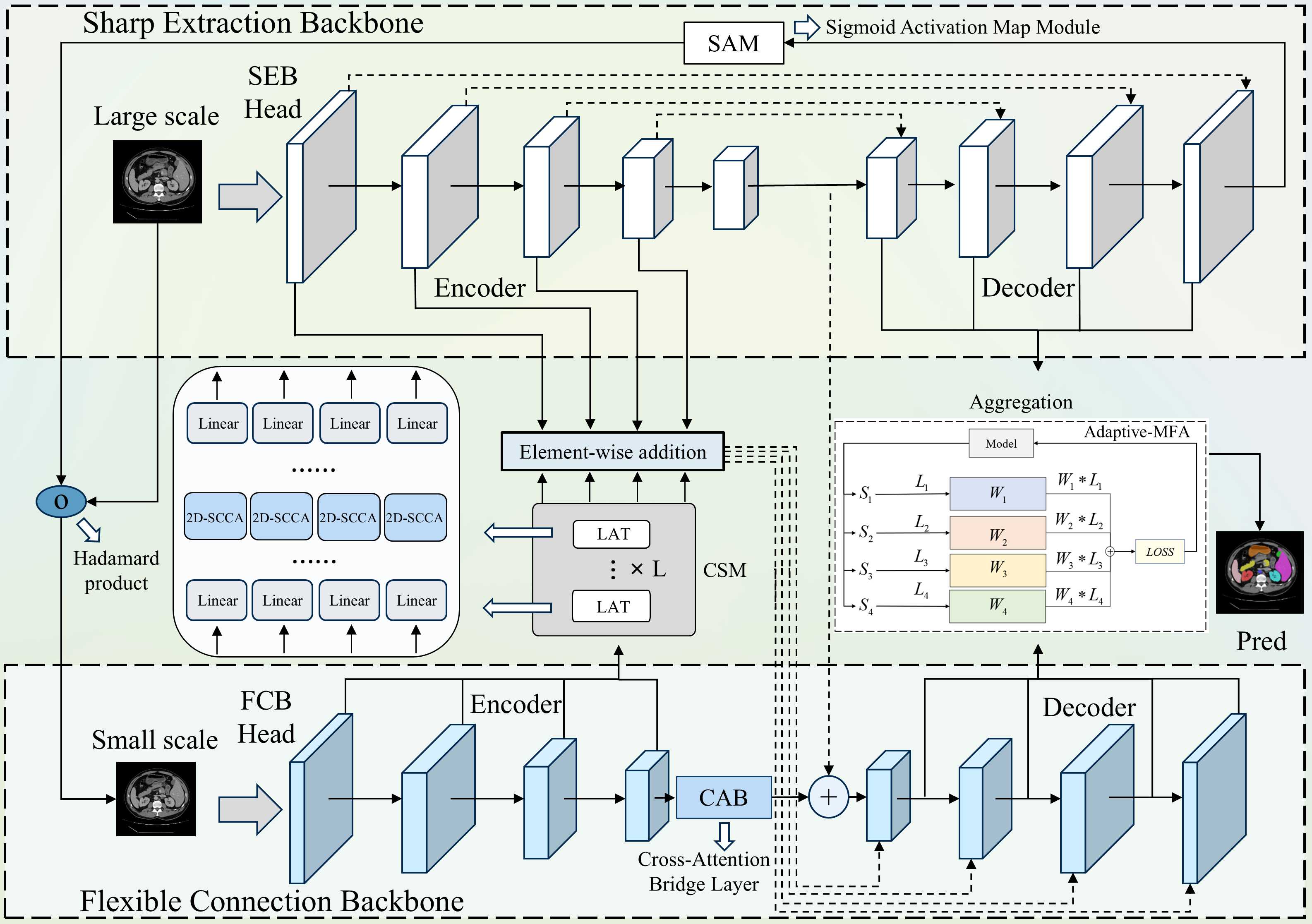}}
\caption{Illustration of the overall MCFNet architecture. The network is built upon two complementary backbones: the Sharp Extraction Backbone for fine details and the Flexible Connection Backbone for global semantics, which are integrated via a Cascaded Skip-connection Module and an Aggregation Module.
}
\label{MCFNet}
\end{figure*}

\subsection{Cascaded Networks for Medical Image Segmentation}
In recent years, cascade networks have become prominent in medical image segmentation. These networks concatenate multiple modules to handle specific subtasks, improving segmentation performance gradually. By integrating feature information at different scales and decomposing tasks into multiple stages, cascade networks optimize segmentation through mechanisms like skip connections and attention mechanisms \cite{ref39}. Experimental results show that cascade networks are effective and often superior for medical image segmentation, with consistent gains in Dice and HD95 across diverse benchmarks (e.g., cardiac structures, and tumor/OARs delineation)\cite{ref40}. Beyond accuracy, cascaded designs improve robustness to domain shifts and small, low-contrast targets, translating into more reliable auto-contouring and reduced manual correction time in clinical workflows—thereby offering tangible benefits for diagnostics, radiotherapy planning, and longitudinal treatment assessment\cite{ref41,ref42}.

The CASCADE model introduces an attention-based decoder that captures local pixel-wise dependencies, addressing a known limitation of Transformer architectures. G-CASCADE~\cite{ref43} employs graph-convolution blocks to refine hierarchical Transformer encoder features, improving segmentation accuracy and robustness. The Cascaded MERIT \cite{ref38} model improves generalization by employing a multiscale self-attention mechanism, which better captures multiscale information than traditional single-scale attention mechanisms. MCANet~\cite{ref44} employs multi-scale cross-axis attention to model lesions and organs with heterogeneous sizes and shapes, strengthening global context and enabling precise target segmentation. \bigskip

Distinguished from existing studies, our method offers enhanced multi-scale modeling, demonstrating leading performance and generalization in robust multi-dataset evaluations.

\begin{figure}[!t]
\centerline{\includegraphics[width=1.0\linewidth]{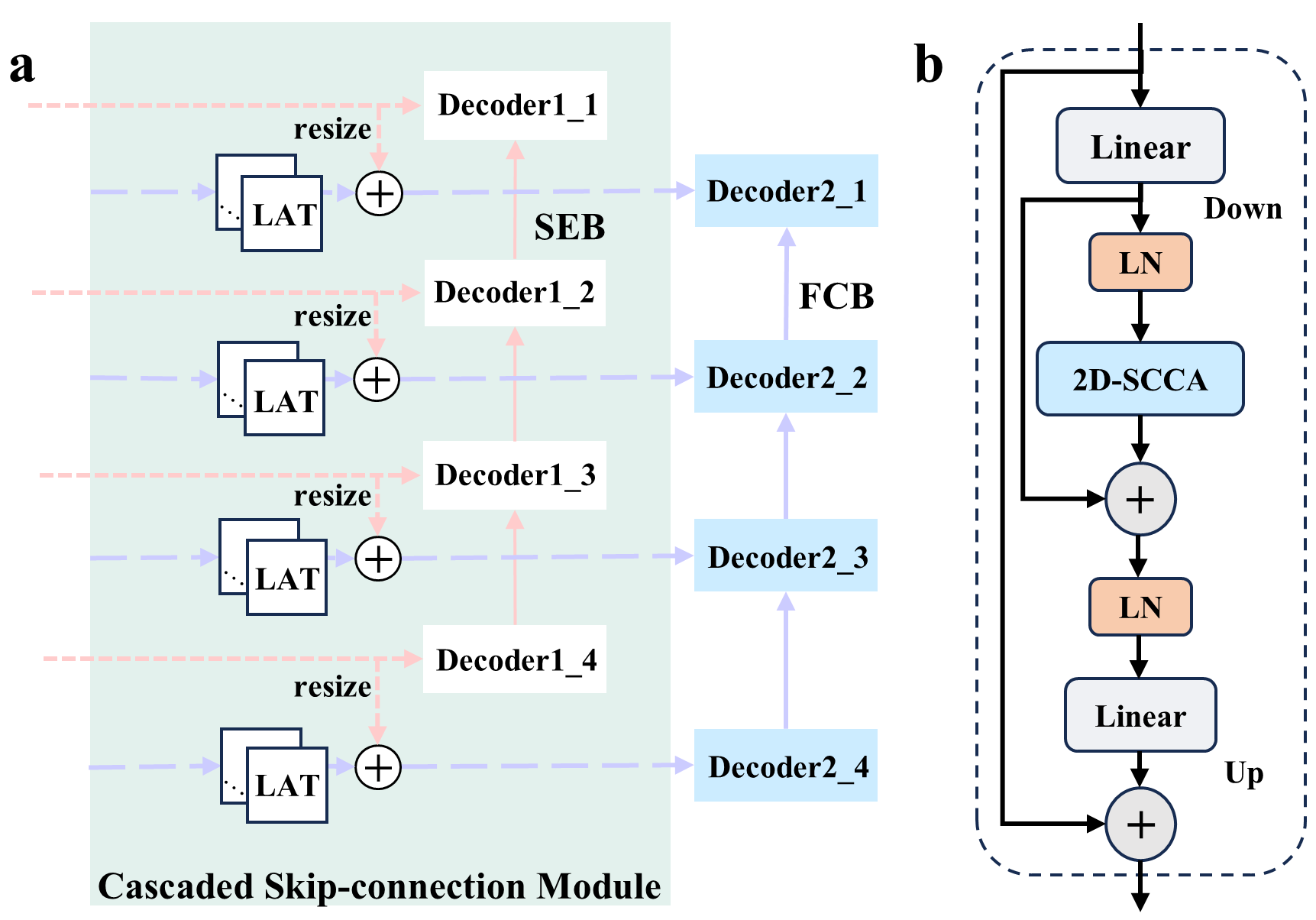}}
\caption{(a) Illustration of the Cascaded Skip-connection Module (CSM). (b)
Illustration of the Linear Attention Transformer Block (LAT) in the CSM.}
\label{CSM}
\end{figure}

\section{Proposed Method}
\label{sec:proposed method}
In this section, we introduce the multi-scale cascade network model MCFNet designed for whole-body OARs segmentation. First, we provide an overall introduction to the model, then we describe each component of the model in detail, and finally, we explain the adaptive loss aggregation strategy we designed.

\subsection{MCFNet for Whole-body OARs Segmentation}
Fig. \ref{MCFNet} shows the overall architecture of our designed MCFNet network. MCFNet consists of three parts: Cascaded Backbones, Decoders with cascaded skip connections, and Aggregation. The Cascaded Backbones module is composed of Sharp Extraction Backbone (SEB) and Flexible Connection Backbone (FCB), enabling multi-scale and multi-resolution input for the model. The Decoders with cascaded skip connections module mainly achieves feature aggregation at skip connections and bridge layers in the cascaded network. The Aggregation module primarily reflects the aggregation of the decoder output layers and the application of our designed adaptive loss aggregation strategy.

\subsection{Cascaded Backbones in MCFNet}
The cascaded backbone network consists of SEB and FCB, responsible for extracting features from images of different resolutions. High-resolution images provide rich details but increase computational complexity. Low-resolution images are more robust to noise but lose details. We use FCB for $224\times224$ images and SEB for $256\times256$ images. FCB uses max pooling and convolution in the encoder, but the downsampling stage can further enhance feature capture. SEB focuses on improving feature extraction during downsampling with fewer parameters and computational load.

Specifically, SEB enhances the extraction of detailed features during downsampling for high-resolution images by incorporating SE attention mechanisms in the encoding layers. The images pass through BatchNorm, MaxPool, and convolution operations in the SEB Stem and encoding layers. The final prediction map is produced from the SEB decoder's highest layer, and a Hadamard product is applied with the input image to generate a new output image. This output is resized to the $224\times224$ input required by FCB. Low-resolution images are processed in the FCB through its Stem and encoding layers, with feature maps output via skip-connections and subsequently aggregated with those from the SEB.

\subsection{Cascaded Skip-connection Module in MCFNet}
As shown in Fig. \ref{CSM}, this section introduces the cascaded skip-connection mechanism. The efficient FCB backbone ensures effective feature extraction, while the lightweight SEB captures detailed information in high-resolution images to complement FCB's limitations.

Specifically, when the encoding layers of FCB reach the lowest level after downsampling, they enter the Cross-Attention Bridge Layer proposed in our previous work, which compensates for the loss of low-level features and aggregates them with the output from SEB's fourth encoding layer. FCB's skip connections first pass through the Linear Attention Transformer Block (LAT), which enables efficient feature extraction across both channel and spatial dimensions. The LAT architecture is based on the parallel attention transformer module (PAT) from our previous work FSCA-Net, and has been improved to accelerate convergence and reduce the number of parameters. After the input tensor passes through the LAT module, it first undergoes dimensionality reduction via a linear layer to align the tensor dimensions with the channel numbers, reducing computational and storage demands. The compressed feature maps then pass through LN, 2D-SCCA attention, and another LN layer, ultimately being restored to their original dimensions through a linear layer.

The feature maps from LAT are aggregated with those from resized SEB and passed into FCB’s decoding layer via skip connections, preserving shallow feature details. In SEB, the highest decoding layer's output ($S_{output}$) is processed by the SAM module, which uses 1x1 convolution and the $Sigmoid$ activation function to enhance key features and suppress less important ones. The calculation formula for the SAM module is as follows: 
\begin{equation}
    Y(i,j) = w\cdot X(i,j) + b,
\end{equation}
\begin{equation}
    Z(i,j) = \sigma(Y(i,j))=\frac{1}{1+{e}^{-Y(i,j)}},
\end{equation}
where the weight is denoted as $W$, the input image is $X$ (with dimensions $H\times W$), the output of the convolution operation is $Y$ (with dimensions $H\times W$), the bias term is denoted as $b$, the output of the Sigmoid activation function is $Z$ (with dimensions $H\times W$), and the values of the output image $Z$ are constrained between 0 and 1. After the $S_{output}$ undergoes the SAM operation, the final output image $Z$ is obtained. This output image is then subjected to Hadamard product operation with the input image $S_{input}$ of SEB, followed by a resize operation, as shown in the following formula:
\begin{equation}
    \hat{O}=S_{input}(i,j)\cdot Z(i,j),
\end{equation}
where $\hat{O}$ is the output image obtained from the Hadamard product. The resize operation is implemented using bilinear interpolation. The bilinear interpolation is defined as follows:
\begin{equation}
\begin{split}
    f(x,y)=& (1-\alpha)\cdot (1-\beta)\cdot pixel(i_1,j_1)\\
    & +\alpha \cdot (1-\beta)\cdot pixel(i_2,j_2)\\
    & +(1-\alpha)\cdot \beta \cdot pixel(i_3,j_3)\\
    & +\alpha \cdot \beta \cdot pixel(i_4,j_4),
    \end{split}
\end{equation}
where the target image coordinates are denoted as $(x, y)$, the original image coordinates are $(i, j)$, and the four nearest neighbor pixels in the original image are $(i_1, j_1)$, $(i_2, j_2)$, $(i_3, j_3)$, $(i_4, j_4)$. Additionally, $\alpha=x-i_1$, $\beta=y-j_1$, and $pixel(i, j)$ represent the pixel value at coordinates $(i, j)$ in the original image. Each decoding layer of SEB and FCB generates a prediction map at different scales, reflecting varying feature levels. These maps are input into the Aggregation module to combine multi-scale information, enhancing the accuracy and robustness of the final predictions.

\begin{figure}[!t]
\centerline{\includegraphics[width=1.0\linewidth]{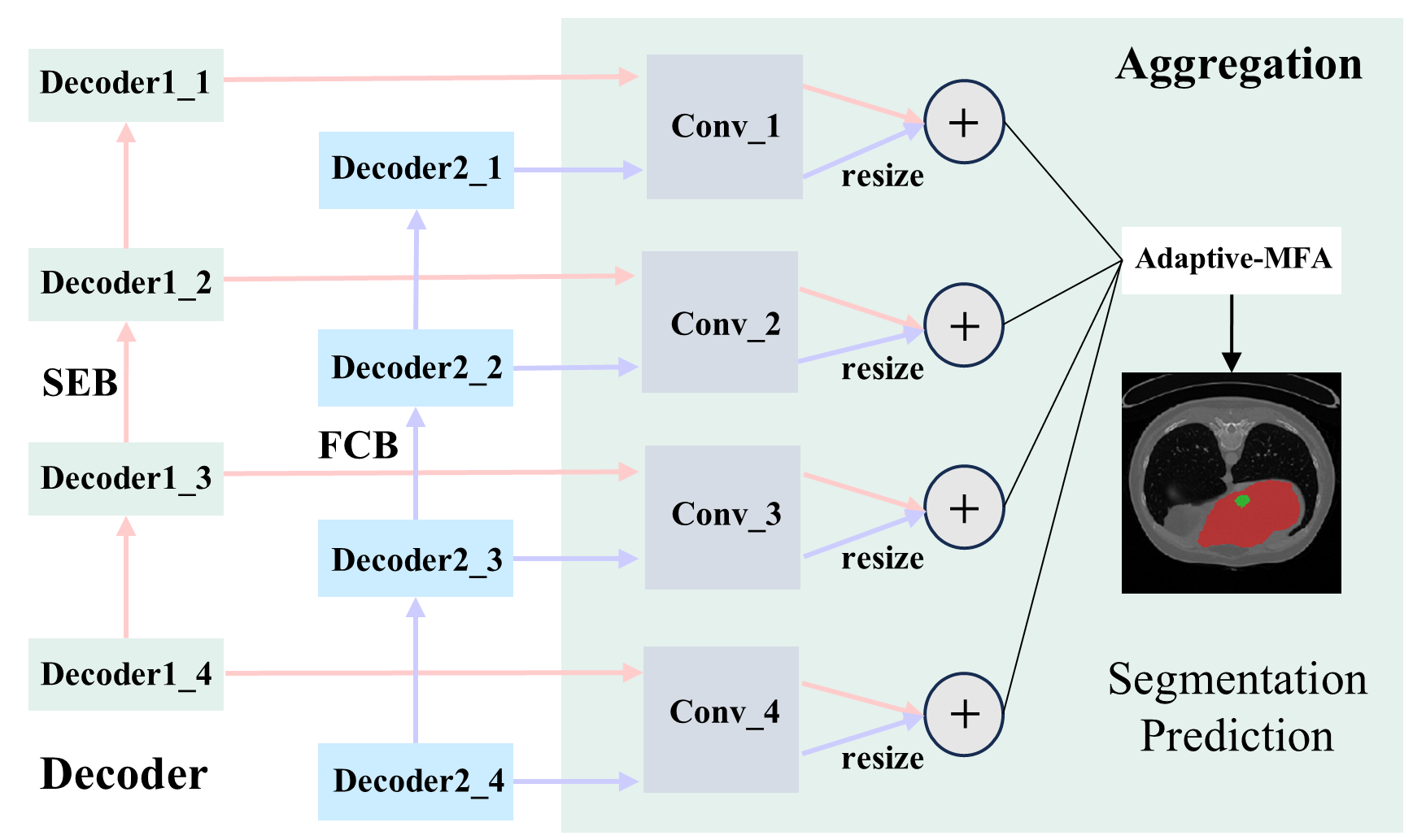}}
\caption{Illustration of the Aggregation module that merges hierarchical multi-scale features from FCB and SEB into a compact representation supervised by Adaptive-MFA.}
\label{Aggregation}
\end{figure}

\subsection{Aggregation in MCFNet}
As shown in Fig. \ref{Aggregation}, the module summarizes the process of multi-scale feature aggregation in the decoding stage and the application of our adaptive loss aggregation strategy. Multi-scale feature aggregation captures global context and fine details, improving segmentation of small, complex structures (e.g., blood vessels and tumors). By merging features from different scales, the model can handle noise, image deformations, and objects of varying sizes, ensuring more stable performance across diverse medical images.

Specifically, the predicted maps output from the four layers of the encoding stage of SEB and FCB are pairwise fed into the corresponding convolutional blocks for channel-wise processing, obtaining feature maps with the desired specified number of channels. The calculation formula for this part is shown below:
\begin{equation}
    O_i=Conv_{(C_i,N,K)}(E_i)(i=1,2,3,4),
\end{equation}
where $E$ is the input feature map, $O$ is the output feature map, $C$ is the number of input channels, $N$ is the number of label classes, and $K$ is the convolution kernel size, with $C_1=64$, $C_2=64$, $C_3=128$, and $C_4=256$. After convolution, bilinear interpolation is used to resize feature maps from FCB and SEB for aggregation. The four scales of feature maps are then processed by an adaptive loss aggregation strategy for data augmentation, optimizing the training. The final output prediction map $Pred$ is obtained by aggregating these layers through adaptive search optimization during training. The aggregation of the four layers of feature maps can be expressed by the following mathematical formula:
\begin{equation}
    Pred=u\times p_1+v\times p_2+w\times p_3+x\times p_4,
\end{equation}
where $p_1$, $p_2$, $p_3$, $p_4$ are the feature maps from the four prediction heads, with $u$, $v$, $w$, $x$ as their corresponding weight coefficients, all set to 1 in our experiments. 

Our model produces accurate segmentation maps. Within the aggregation module, combining multi-scale features refines boundaries and improves localization, thereby enhancing overall prediction performance and clinical utility.

\begin{figure}[!t]
\centering
\includegraphics[width=1.0\linewidth]{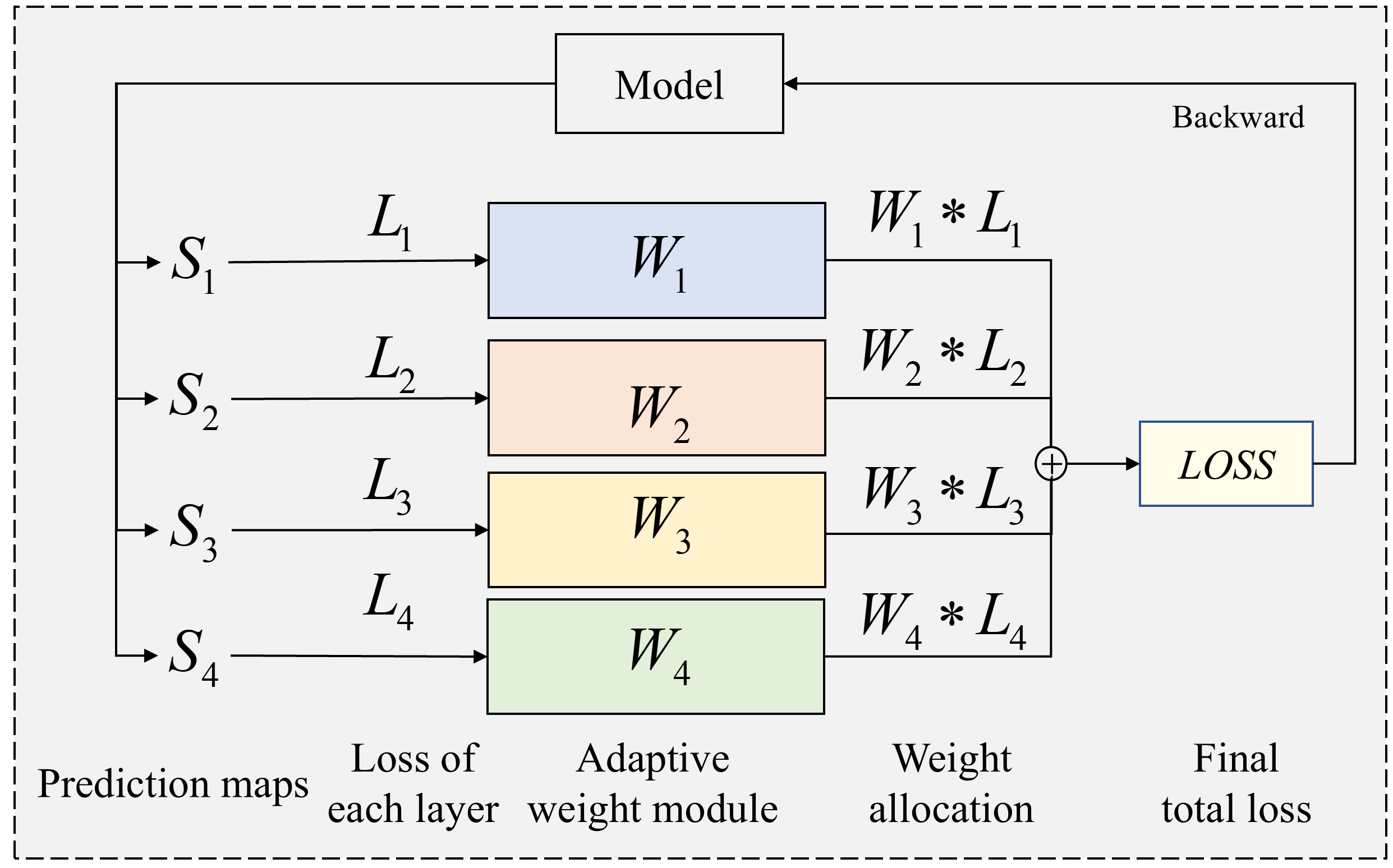}
\caption{Flowchart of the Adaptive-MFA Strategy, which is executed concurrently with the training process.}
\label{Adaptive-MFA}
\end{figure}

\subsection{Adaptive-MFA: Adaptive Multi-scale Feature-Mixing Loss Aggregation}
To optimize model training, we design an adaptive loss aggregation strategy, primarily by combining different prediction maps to create new ones. The adaptive selection and weighting of prediction maps help optimize the final results for better prediction accuracy in varying scenarios.

The strategy flowchart in Fig. \ref{Adaptive-MFA} shows how we combine output prediction maps $P_1$, $P_2$, $P_3$, and $P_4$ from the decoder’s four layers, forming 15 non-empty subset combinations. These are divided into four sets $S_1$, $S_2$, $S_3$, with adaptive weights $W_1$, $W_2$, $W_3$, and $W_4$ adjusted during training. We remove the constraint $W_1 + W_2 + W_3 + W_4 = 1$ because it hinders prediction maps with minimal contribution from playing an important role in model training. Initially, all weights are set equal: $W_1 = W_2 = W_3 = W_4$. The specific contents of the four sets are as follows:
\begin{equation}
    S_1 = \Big\{\{p_1\},\{p_2\},\{p_3\},\{p_4\}\Big\}, 
\end{equation}
    
\begin{equation}
\begin{split}
   S_2 = &\Big\{\{p_1,p_2\},\{p_1,p_3\},\{p_1,p_4\},\\
    & \{p_2,p_3\},\{p_2,p_4\},\{p_3,p_4\}\Big\}, 
\end{split}
\end{equation}

\begin{equation}
\begin{split}
    S_3 = &\Big\{\{p_1,p_2,p_3\},\{p_1,p_2,p_4\},\\
    &\{p_1,p_3,p_4\},\{p_2,p_3,p_4\}\Big\},\\
\end{split}
\end{equation}

\begin{equation}
    S_4 = \Big\{\{p_1,p_2,p_3,p_4\}\Big\}.
\end{equation}

We perform feature map aggregation operations according to the combinations of each subset within each set. Subsequently, we calculate the total loss for each of the four sets, denoted as $L_1$, $L_2$, $L_3$ and $L_4$. By incorporating the corresponding weight coefficients, we obtain the final total loss, $LOSS$, which is expressed as:
\begin{equation}
LOSS = W_1*L_1 + W_2*L_2 + W_3*L_3 + W_4*L_4,
\end{equation}
the $LOSS$ obtained after each epoch during the training phase is fed back to the model itself, optimizing the model training and adaptively adjusting the weights. This process continues throughout the entire model training phase.

In summary, randomly combining and aggregating the outputs of the four-layer decoder allows full utilization of multi-level features, enhancing the model's predictive capability, robustness, and flexibility. This approach also provides new ideas and methods for model design and optimization. Furthermore, through experiments, we find that the weight coefficients of prediction maps tend to favor different sets across different datasets. Therefore, adaptively searching for the optimal weight coefficients better meets the demands of different scenarios, thereby improving the model's generalization ability and prediction accuracy. This adaptive search method dynamically adjusts weight distribution based on the characteristics of different datasets, enabling excellent performance across various scenarios.

\section{Experiments}
\label{sec:experiments}
This section introduces the datasets and implementation settings, then reports detailed quantitative and qualitative results, followed by thorough ablation studies and discussion of the proposed method.

\subsection{Datasets}
To assess whole-body OARs and lesion segmentation, we evaluate MCFNet on ten datasets spanning multiple anatomical regions and imaging protocols. For each dataset, we report the source, data volume, and annotation scope.

\subsubsection{\textbf{HaN (2015)}} This public dataset contains CT and MRI images of patients' head and neck, used for image-guided radiotherapy planning. The dataset provides segmentation labels for six regions: brainstem, mandible, left and right parotid gland, left and right submandibular gland. After preprocessing, we merged these labels into a single file. The dataset consists of images from 17 tumor patients, with 14 cases used for training (1924 slices) and 3 cases for testing.

\subsubsection{\textbf{SegTHOR (2019)}}This CT dataset, part of the ISBI 2019 challenge, is designed for thoracic risk organ segmentation, including the heart, trachea, aorta, and esophagus. These organs differ in spatial and appearance characteristics. After preprocessing, we divided the dataset of 40 cases into 32 for training (6035 slices) and 8 for testing.

\subsubsection{\textbf{CHAOS (2019)}} This multimodal abdominal segmentation dataset comprises paired CT–MRI volumes with corresponding annotations. MRI data includes T1-DUAL and T2-SPIR sequences, with annotations for liver, spleen, left kidney, and right kidney. After preprocessing, we split the data of 40 cases into 32 for training (1026 slices) and 8 for testing.

\subsubsection{\textbf{LiTS (2017)}} This dataset focuses on liver and tumor segmentation in CT images. It was collected from seven medical centers and included in the ISBI 2017, MICCAI 2017, and MICCAI 2018 challenges. From 131 patients, we randomly select data from 40, with 32 cases for training (4599 slices) and 8 cases for testing.

\subsubsection{\textbf{KiTS (2019)}} This dataset focuses on kidney and tumor segmentation in CT images and was part of the MICCAI 2019 challenge. It includes data from 300 kidney cancer patients who underwent nephrectomy. From 210 available cases, we randomly select 80, with 64 cases for training (4754 slices) and 16 cases for testing.

\subsubsection{\textbf{KiTS (2023)}} This dataset extends the KiTS (2019) dataset to include cyst segmentation, increasing the total cases from 210 to 489. It was part of the MICCAI 2023 challenge. From 489 cases, we randomly select 80, with 64 cases for training (4968 slices) and 16 cases for testing.

\subsubsection{\textbf{LUNG (2018)}} This dataset is part of the MSD Medical Image Segmentation Decathlon Challenge, focusing on lung tumor segmentation from CT scans. It includes thin-slice CT scans from 63 non-small cell lung cancer patients. After preprocessing, we divide it into 50 cases for training (1264 slices) and 13 cases for testing.

\begin{figure*}[!t]
    \centering
    \captionsetup[subfigure]{justification=centering}
    \subfloat[CHAOS]{\includegraphics[width=0.45\linewidth]{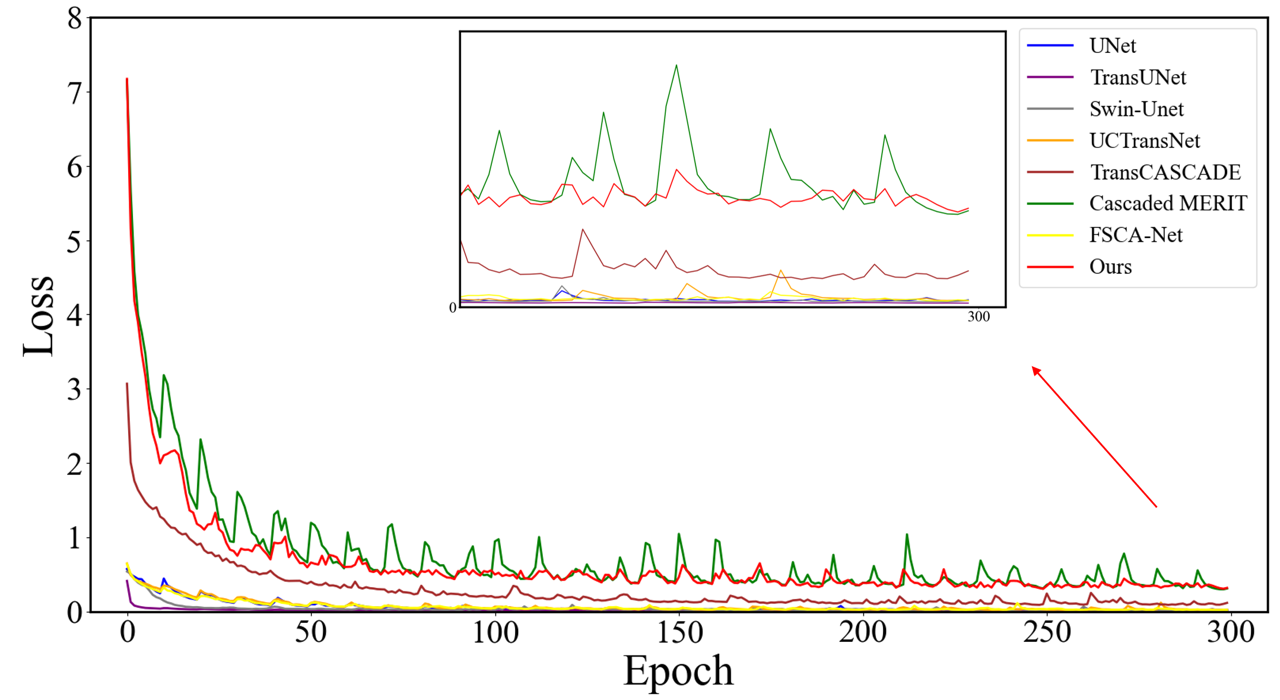}}
    \hfil
    \subfloat[LUNG]{\includegraphics[width=0.45\linewidth]{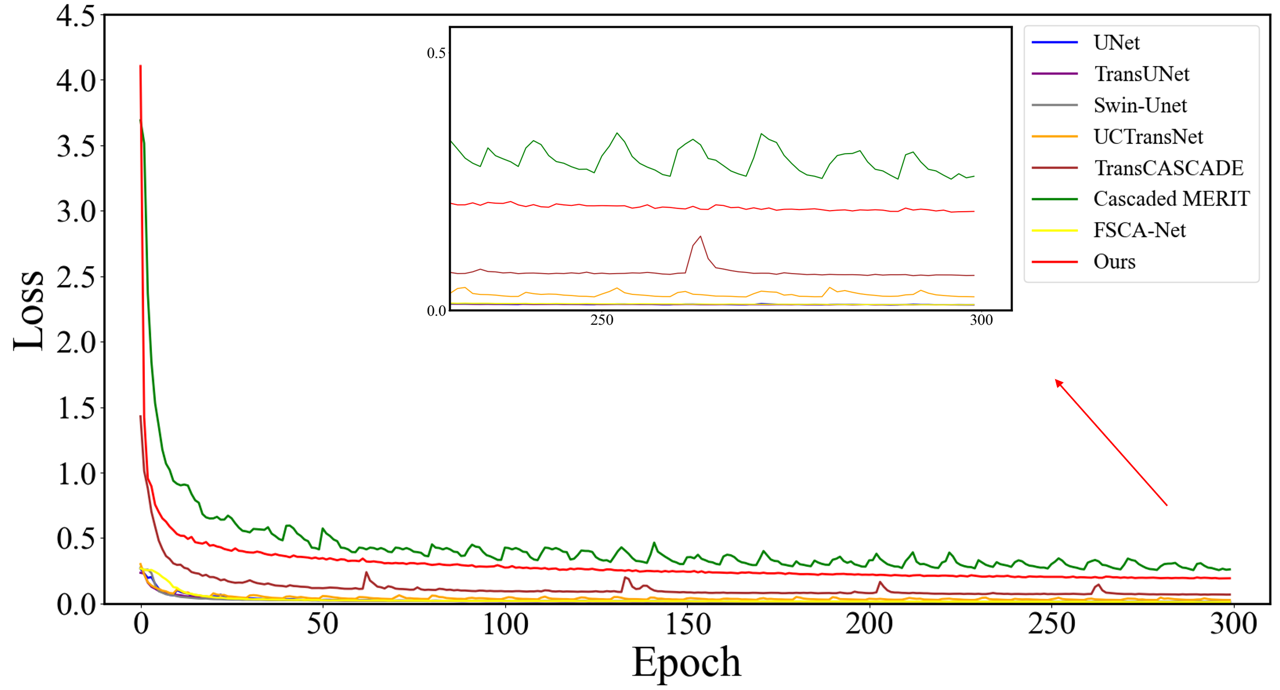}}
    \hfill
    \subfloat[CPCGEA]{\includegraphics[width=0.45\linewidth]{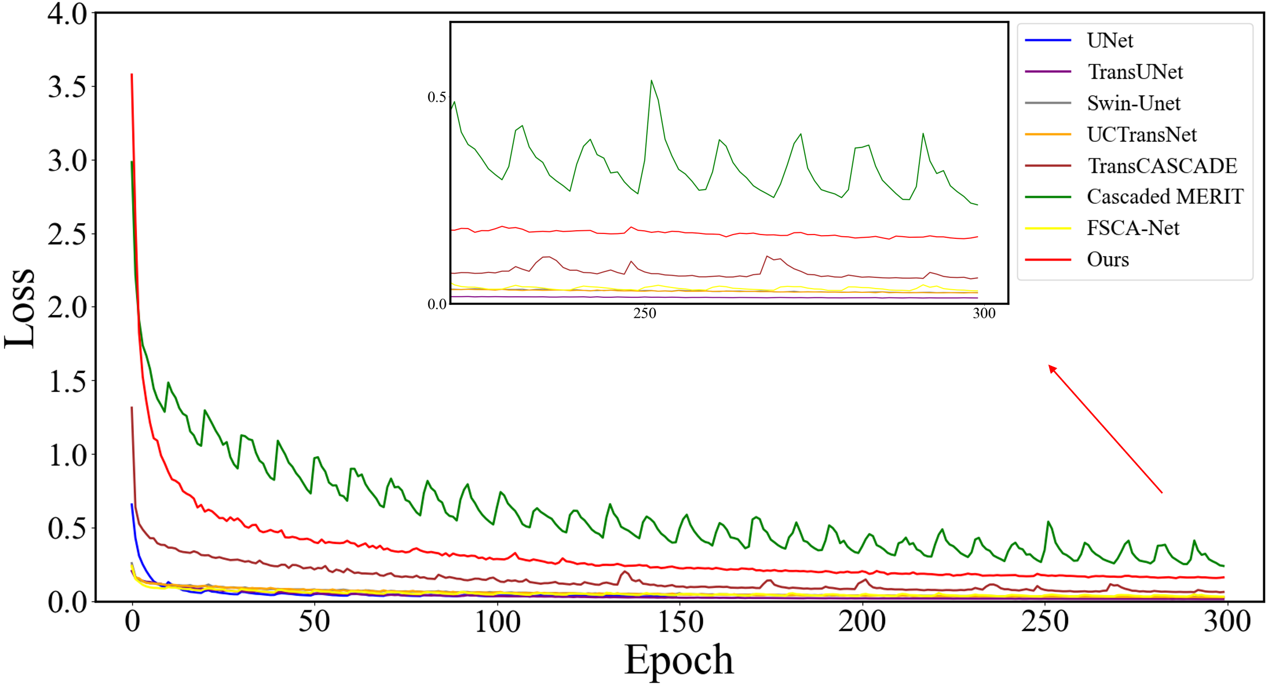}}
    \hfil
    \subfloat[KNEE]{\includegraphics[width=0.45\linewidth]{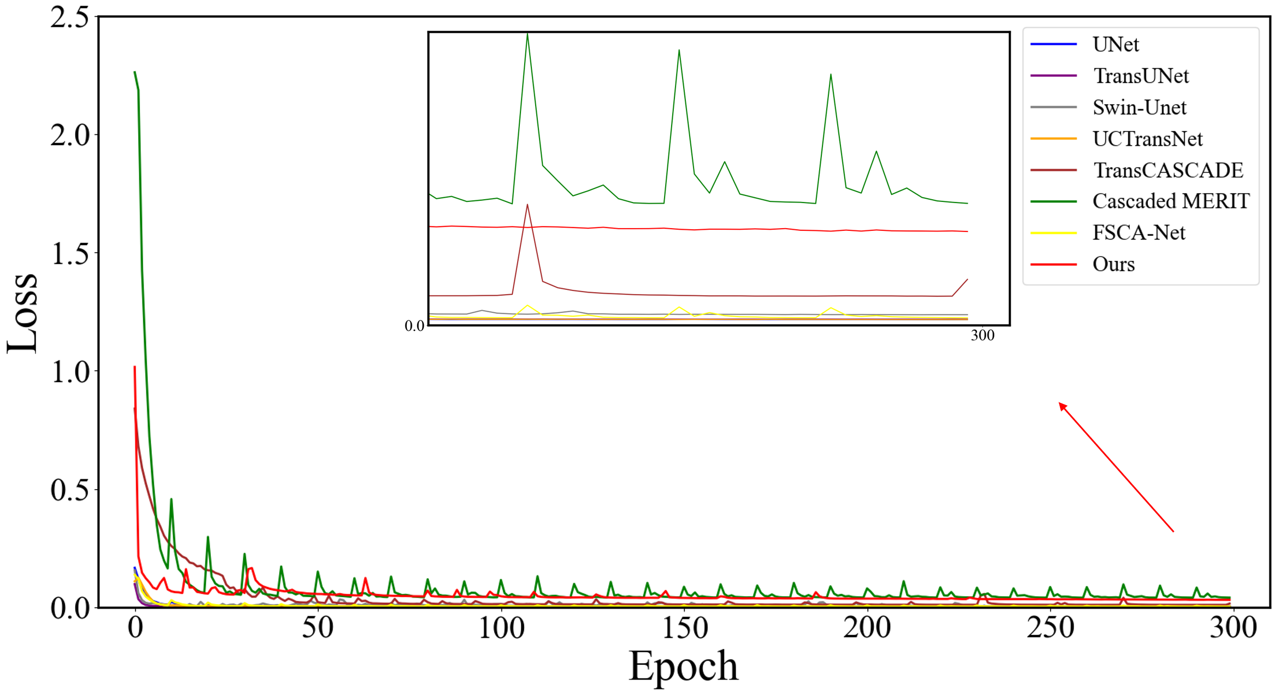}}
    \caption{Training loss curves of the proposed method and seven comparison methods across four datasets. The stable downward trend of the loss demonstrates the orderly progression of model training.}
    \label{loss}
\end{figure*}

\subsubsection{\textbf{CPCGEA (2023)}} This dataset contains MRI data of prostate cancer patients, including DWI and T2WI sequences. The images are affected by poor clarity and noise, with artifacts like streaks and shadows due to image acquisition issues. It includes 172 cases with annotated prostate cancer regions. We randomly split the data into 139 training cases (832 slices) and 33 test cases.

\subsubsection{\textbf{KNEE (2024)}} This dataset consists of MRI images of patients' leg bones, focusing on femur segmentation. It includes data from 59 patients. After preprocessing, we use 47 cases for training (8171 slices) and 12 cases for testing.

\subsubsection{\textbf{AutoPET (2023)}} AutoPET is a large-scale PET/CT dataset for whole-body tumor segmentation, included in the MICCAI 2022 and MICCAI 2023 challenges. It contains 1014 paired PET-CT images from 900 patients, focusing on malignant melanoma, lymphoma, and lung cancer. The dataset includes three-dimensional FDG-PET and CT images, with manually annotated tumor masks. We select PET images and annotations from 80 patients, dividing the data into 64 training cases and 16 test cases, with 2558 slices in the training set.

\subsection{Evaluation Metrics}
We evaluate segmentation models using appropriate metrics based on the dataset type.

For multi-class segmentation, we evaluate using the Dice similarity coefficient (DSC) and the 95th-percentile Hausdorff distance (HD95). DSC measures overlap between the prediction and ground truth (higher is better), while HD95 quantifies boundary error at the 95th percentile (lower is better).

For single-class segmentation, we report DSC, HD95, Recall (Sensitivity), and Precision (Positive Predictive Value).  Recall evaluates the model’s ability to capture all target voxels (penalizing false negatives), and Precision measures the proportion of predicted positives that are correct (penalizing false positives). Together, these metrics provide a balanced evaluation of volumetric agreement and boundary accuracy.

\begin{figure*}[!t]
\centerline{\includegraphics[width=1.0\linewidth]{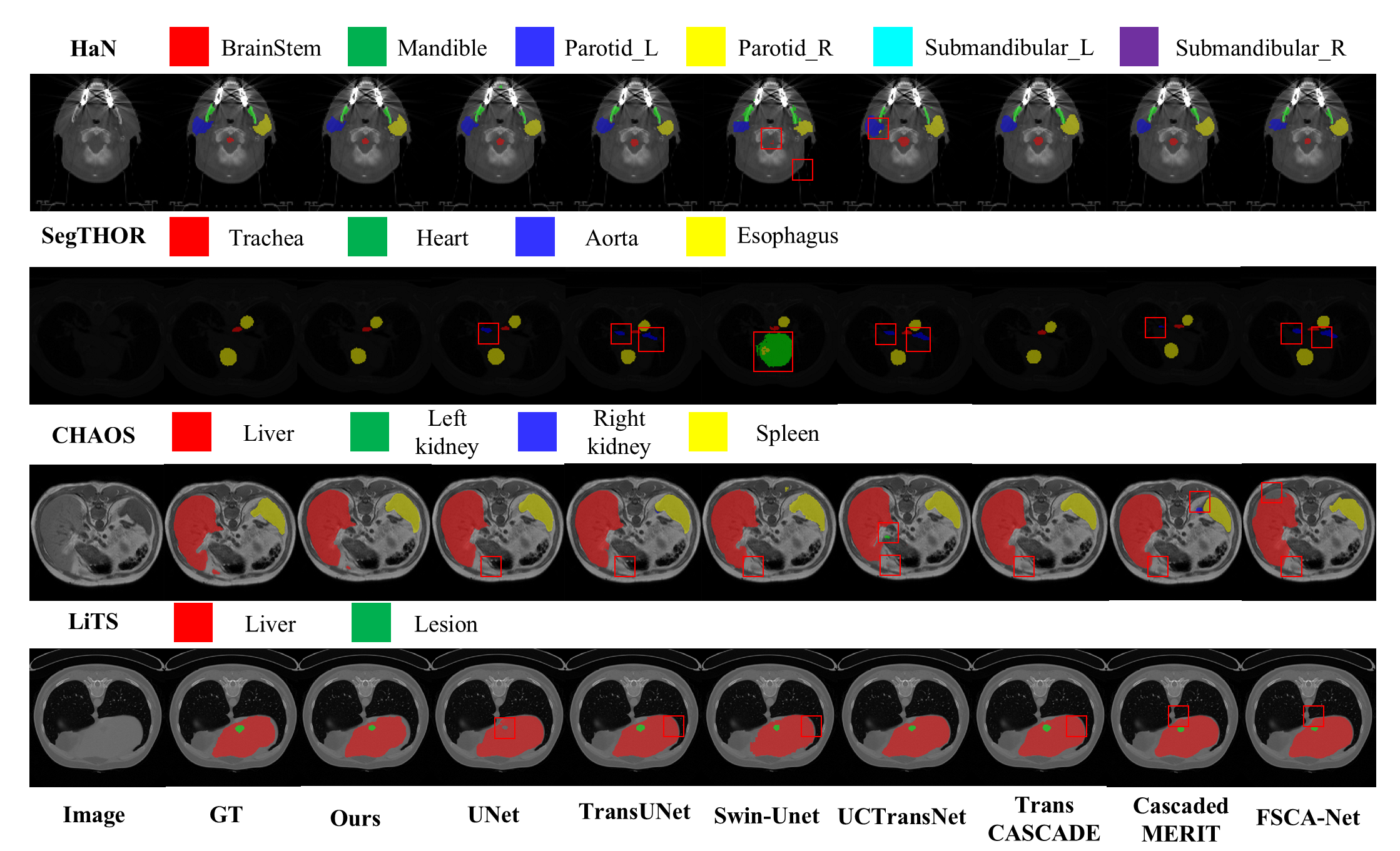}}
\caption{Visual examples of segmentation on the HaN, SegTHOR, CHAOS, and LiTS datasets. In the image, different organs or lesions are highlighted using distinct color annotations.}
\label{fourDatasets}
\end{figure*}

\subsection{Implementation Detail}
All experiments are conducted on a NVIDIA A6000 GPU. We standardize the resolution of all input images to $256\times 256$. The max training epoch is 300 with a batch size of 16 for all experiments. We use the Adam optimizer for model training with an initial learning rate of 0.001 and a cosine annealing learning rate decay strategy. The weight decay is set to 0.0001. The loss function used is a combination of Dice Loss and Cross-Entropy Loss. It can be defined as:
\begin{equation}
\begin{split}
L(Y,P) =& 1 - \sum_{i=1}^I(\lambda \frac{2 * \sum_{n=1}^N {Y}_{n,i}\cdot {P}_{n,i}}{\sum_{n=1}^N {Y}_{n,i}^2 + \sum_{n=1}^N {P}_{n,i}^2}  \\
& + \sum_{n=1}^N {Y}_{n,i}log{P}_{n,i}),
\label{eq}
\end{split}
\end{equation}
$I$ is the number of classes, $N$is the total number of voxel. ${Y}_{n,i}$ and ${P}_{n,i}$  are the ground truth and output probability for the $i$-th class at voxel $n$, respectively.

\begin{table*}[!t]
    \centering
    \captionsetup{justification=centering}
    \caption{COMPARISONS WITH STATE-OF-THE-ART MODELS ON THE HAN DATASET.}
    \label{tab:HAN}
    \resizebox{\textwidth}{!}{%
    \begin{tabular}{@{}ccccccccc@{}}
    \toprule
        \multirow{2}*{Method}& DSC\textcolor{red}{$\uparrow$}& HD95\textcolor{red}{$\downarrow$}& BrainStem& Mandible& Parotid$_L$& Parotid$_R$& Submandibular$_L$& Submandibular$_R$\\ 
        ~ & (\%, mean)& (mm, mean)&  \multicolumn{6}{c}{DSC$\textcolor{red}{\uparrow}$(\%)}\\ \midrule
        \text { UNet } & 72.98 & 3.260 & 77.44 & 89.45 & 73.68 & 76.09 & 59.08 & 62.17 \\
        \text { TransUNet } & 71.48 & 5.412 & 76.12 & 90.81 & 71.97 & 71.16 & 56.96 & 61.82 \\
        \text { Swin-Unet } & 68.75 & 7.541 & 73.51 & 91.67 & 70.53 & 73.27 & 50.63 & 52.85 \\
        \text { UCTransNet } & 70.36 & 4.328 & 77.26 & 88.85 & 70.38 & 71.03 & 61.23 & 53.43 \\
        \text { TransCASCADE } & 71.56 & 3.184 & 79.39 & 87.67 & 70.75 & 75.20 & 51.44 & 64.89 \\
        \text { Cascaded MERIT } & 74.71 & 2.532 & 79.17 & 92.14 & 75.70 & 78.57 & 54.78 & 67.89 \\
        \text { FSCA-Net } & 73.31 & 4.822 & 75.52 & 90.19 & 73.62 & 72.36 & 64.06 & 64.11 \\\midrule
        \textbf {Ours} & \textbf{77.52} & \textbf{2.453} & \textbf{79.52} & \textbf{93.47} & \textbf{77.59} & \textbf{78.61} & \textbf{68.02} & \textbf{67.92} \\
    \bottomrule
    \end{tabular}
    }
\end{table*}

\subsection{Comparsion With Other Methods}
We validate the performance of MCFNet across ten datasets using seven state-of-the-art methods: U-Net, TransUNet, Swin-Unet, UCTransNet, TransCASCADE, Cascaded MERIT, and FSCA-Net. Fig. \ref{loss} shows the training loss trends of all methods across four datasets, while Fig. 8, 9, 10, and 11 display visual segmentation examples for each dataset, with errors highlighted in red boxes. The specific qualitative and quantitative analysis results are as follows:

\subsubsection{Experiments on HaN} Our proposed method and seven comparison models segment six regions in this dataset. As shown in the Table \ref{tab:HAN}, the Cascaded MERIT model achieves excellent segmentation performance, with an average DSC of 74.71\% and an average HD95 of 2.532mm. Our designed MCFNet improves the average DSC by 2.81\% and reduces the average HD95 by 0.079mm compared to Cascaded MERIT. The results in Table \ref{tab:HAN} indicate that our model achieves a substantial performance gain on the left submandibular gland.

In the first row of Fig. \ref{fourDatasets}, we see that our method avoids both under-segmentation and over-segmentation. However, UNet misidentifies the mandibular region, Swin-Unet under-segments the brainstem, and UCTransNet struggles to distinguish between the left and right parotid glands.

\subsubsection{Experiments on SegTHOR} Our proposed method and seven comparison models segment the trachea, heart, aorta, and esophagus. As shown in Table \ref{tab:SEGTHOR}, state-of-the-art methods like TransUNet, UCTransNet, and FSCA-Net achieve over 80\% DSC and reduce the average HD95 to below 5mm. Our method achieves an average DSC of 83.27\% and HD95 of 4.187mm, outperforming others in segmenting these organs.

As shown in the second row of Fig. \ref{fourDatasets}, our designed model accurately identifies and segments the risk organs. However, UNet, TransUNet, UCTransNet, Cascaded MERIT, and FSCA-Net all mistakenly identify the background as the aorta. Additionally, Swin-Unet incorrectly identifies the esophagus region as the heart.

\subsubsection{Experiments on CHAOS} The comparison results are shown in Table \ref{tab:CHAOS}. We can see that the cascade networks TransCASCADE \cite{ref40} and Cascaded MERIT achieve relatively good segmentation results for the four abdominal organs, with average DSCs of 90.26\% and 90.69\%, and average HD95s of 3.695mm and 3.914mm, respectively. The traditional TransUNet achieves an average DSC of 91.19\%, making it the best-performing model among the comparisons. Achieving 92.80\% average DSC and 3.676 mm average HD95, the proposed method surpasses all competing models.

As shown in the third row of Fig. \ref{fourDatasets}, most comparison models under-segment the liver, and Cascaded MERIT also struggles with identifying the spleen and right kidney. Our model achieves generally accurate segmentation, but the spleen's edges could still be improved.

\subsubsection{Experiments on LiTS} All results for this dataset are shown in Table \ref{tab:LITS}. We can see that TransCASCADE is the best-performing advanced segmentation method, achieving an average DSC of 72.56\%. For liver tumor segmentation, TransUNet achieves a DSC of 51.93\%, which is currently the best. Our designed model achieves better segmentation results for tumors, with an average DSC of 73.79\% and a DSC of 53.35\% for tumor regions.

\begin{table}[!t]
    \centering
    \captionsetup{justification=centering}
    \caption{COMPARISONS WITH STATE-OF-THE-ART MODELS ON THE SEGTHOR DATASET.}
    \label{tab:SEGTHOR}
    \resizebox{\columnwidth}{!}{%
    \begin{tabular}{@{}ccccccccc@{}}
    \toprule
        \multirow{2}*{Method}& DSC\textcolor{red}{$\uparrow$}& HD95\textcolor{red}{$\downarrow$}& Trachea & Heart & Aorta & Esophagus\\ 
        ~ & (\%, mean)& (mm, mean)&  \multicolumn{4}{c}{DSC\textcolor{red}{$\uparrow$}(\%)}\\ \midrule
        UNet & 75.30 & 6.190 & 56.06 & 84.27 & 82.25 & 78.62 \\
        TransUNet & 80.79 & 4.542 & 64.24 & 88.31 & 84.70 & 85.91 \\
        Swin-Unet & 71.32 & 7.208 & 47.12 & 85.60 & 77.43 & 75.15 \\
        UCTransNet & 80.86 & 5.115 & 64.36 & 87.91 & 85.32 & 85.85 \\
        TransCASCADE & 77.18 & 4.608 & 58.54 & 89.26 & 80.37 & 80.55 \\
        Cascaded MERIT & 78.92 & 4.272 & 59.16 & 89.30 & 83.65 & 83.58 \\
        FSCA-Net & 80.75 & 4.947 & 65.56 & 85.84 & 84.48 & 87.12 \\ \midrule
        \textbf{Ours} & \textbf{83.27} & \textbf{4.187} & \textbf{69.37} & \textbf{89.55} & \textbf{87.75} & \textbf{87.40} \\
    \bottomrule
    \end{tabular}
    }
\end{table}

\begin{table}[!t]
    \centering
    \captionsetup{justification=centering}
    \caption{COMPARISONS WITH STATE-OF-THE-ART MODELS ON THE CHAOS DATASET.}
    \label{tab:CHAOS}
    \resizebox{\columnwidth}{!}{
        \begin{tabular}{ccccccc}
        \toprule
            \multirow{2}*{Method}&  DSC\textcolor{red}{$\uparrow$}& HD95\textcolor{red}{$\downarrow$}& Liver & Kidney$_L$ & Kidney$_R$ & Spleen\\ 
            ~ & (\%, mean)& (mm, mean)&  \multicolumn{4}{c}{DSC\textcolor{red}{$\uparrow$}(\%)}\\ \midrule
            UNet & 90.82 & 8.990 & 94.02 & 90.42 & 90.43 & 88.43 \\
            TransUNet & 91.19 & 5.902 & 94.40 & 91.92 & 91.06 & 87.40 \\
            Swin-Unet & 90.17 & 4.366 & 93.61 & 90.23 & 89.65 & 87.20 \\
            UCTransNet & 88.94 & 6.522 & 91.26 & 90.05 & 86.96 & 87.49 \\
            TransCASCADE & 90.26 & 3.695 & 93.66 & 91.61 & 89.79 & 85.99 \\
            Cascaded MERIT & 90.69 & 3.914 & 93.57 & 91.25 & 91.74 & 86.18 \\
            FSCA-Net & 89.83 & 9.277 & 93.55 & 91.12 & 88.25 & 86.41 \\ \midrule
            \textbf{Ours} & \textbf{92.80} & \textbf{3.676} & \textbf{94.87} & \textbf{93.45} & \textbf{92.05} & \textbf{90.83} \\
        \bottomrule
        \end{tabular}
    }
\end{table}

\begin{table}[!t]
    \centering
    \captionsetup{justification=centering}
    \caption{COMPARISONS WITH STATE-OF-THE-ART MODELS ON THE LITS DATASET.}
    \label{tab:LITS}
    \resizebox{\columnwidth}{!}{%
    \begin{tabular}{ccccc}
    \toprule
        \multirow{2}*{Method}& DSC\textcolor{red}{$\uparrow$}& HD95\textcolor{red}{$\downarrow$}& Liver & Lesion\\ 
        ~ & (\%, mean)& (mm, mean)&  \multicolumn{2}{c}{DSC\textcolor{red}{$\uparrow$}(\%)}\\ \midrule
        UNet & 64.10 & 31.754 & 92.11 & 36.11\\
        TransUNet & 71.46 & 25.951 & 90.97 & 51.93 \\
        Swin-Unet & 66.09 & 30.759 & 91.83 & 40.35 \\
        UCTransNet & 66.77 & 29.473 & 92.66 & 40.87 \\
        TransCASCADE & 72.56 & 25.351 & 94.16 & 50.96 \\
        Cascaded MERIT & 72.36 & 24.521 & 93.25 & 51.47 \\
        FSCA-Net & 68.54 & 23.897 & 93.04 & 44.04 \\ \midrule
        \textbf{Ours} & \textbf{73.79} & \textbf{23.326} & \textbf{94.22} & \textbf{53.35} \\
    \bottomrule
    \end{tabular}
    }
\end{table}

As shown in the fourth row of Fig. \ref{fourDatasets}, our model provides accurate segmentation for the liver and tumors. However, fine delineation of tumor boundaries remains challenging. Models such as UNet often struggle to detect small lesions, and even Cascaded MERIT and FSCA-Net exhibit suboptimal performance in capturing hepatic boundary details.

\subsubsection{Experiments on KiTS (2019)} The dataset targets segmentation of the kidneys and renal tumors. As shown in Table \ref{tab:KITS(2019)}, our designed model achieves an average DSC that is 2.21\% higher than the state-of-the-art model Cascaded MERIT and an average HD95 that is 2.448mm lower. Additionally, MCFNet improves tumor segmentation DSC by 4.49\% over TransCASCADE, and achieves a further 0.21\% DSC gain on kidney segmentation compared with Cascaded MERIT.

\begin{figure*}[!t]
\centerline{\includegraphics[width=1.0\linewidth]{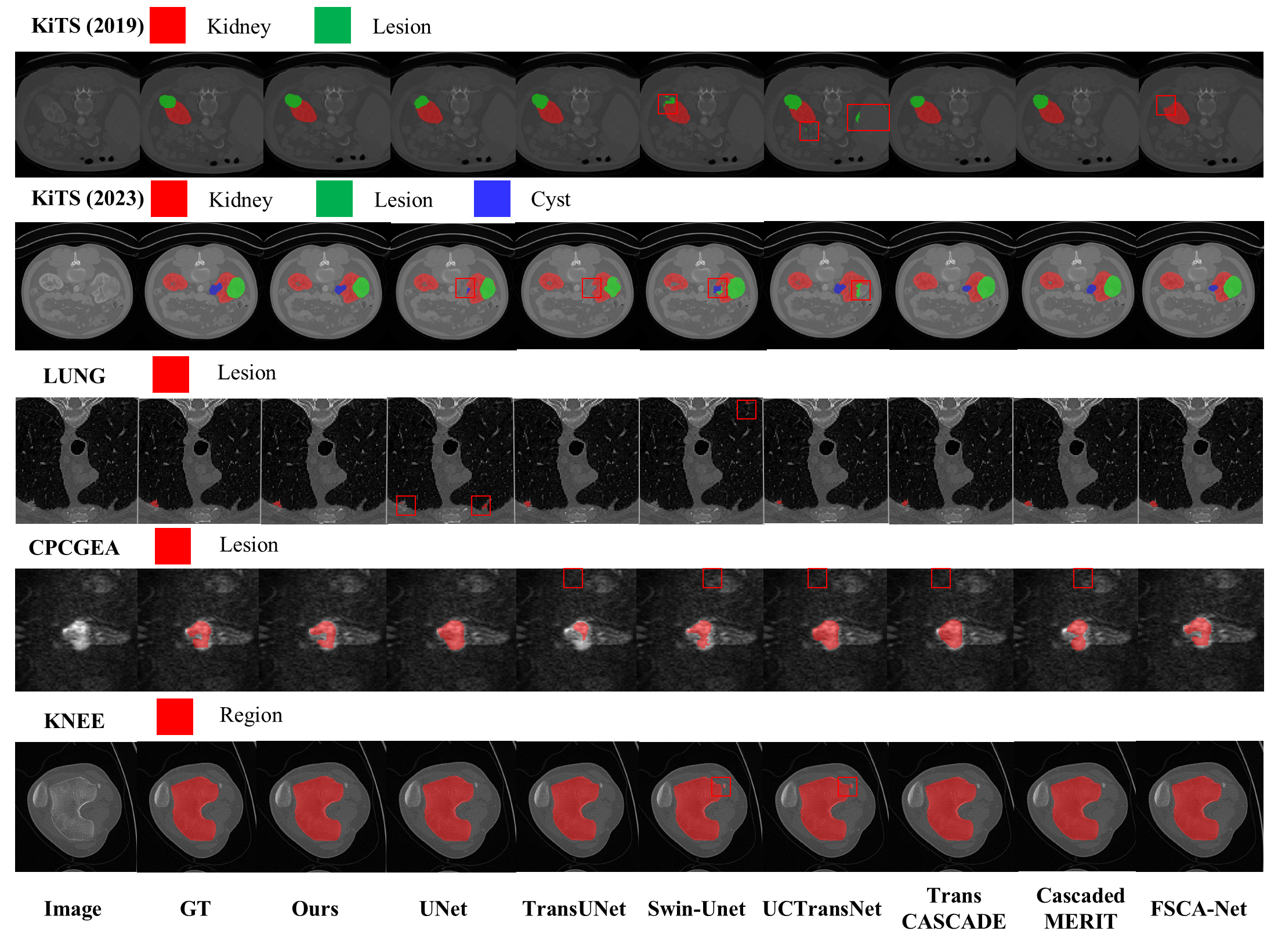}}
\caption{Visual examples of segmentation on the KiTS (2019), KiTS (2023), LUNG, CPCGEA and KNEE datasets. In the image, different organs or lesions are highlighted using distinct color annotations.}
\label{fiveDatasets}
\end{figure*}

As shown in the first row of Fig. \ref{fiveDatasets}, we show the segmentation effects on an image containing both the kidney and tumor. Swin-Unet has scattered segmentation in the tumor region, UCTransNet mistakenly identifies some background areas as tumors, and FSCA-Net fails to recognize the tumor regions.

\begin{table}[!t]
    \centering
    \captionsetup{justification=centering}
    \caption{COMPARISONS WITH STATE-OF-THE-ART MODELS ON THE KITS (2019) DATASET.
}
    \label{tab:KITS(2019)}
    \resizebox{\columnwidth}{!}{%
    \begin{tabular}{ccccc}
    \toprule
        \multirow{2}*{Method}& DSC\textcolor{red}{$\uparrow$}& HD95\textcolor{red}{$\downarrow$}& Kidney & Lesion\\ 
        ~ & (\%, mean)& (mm, mean)&  \multicolumn{2}{c}{DSC\textcolor{red}{$\uparrow$}(\%)}\\ \midrule
        UNet & 65.20 & 23.039 & 87.99 & 42.41\\
        TransUNet & 66.41 & 34.096 & 89.52 & 43.31 \\
        Swin-Unet & 65.21 & 33.455 & 85.09 & 45.33 \\
        UCTransNet & 64.31 & 43.377 & 87.25 & 41.37 \\
        TransCASCADE & 68.14 & 25.023 & 89.78 & 46.43 \\
        Cascaded MERIT & 68.50 & 22.970 & 90.29 & 46.38 \\
        FSCA-Net & 60.59 & 41.617 & 85.97 & 35.22 \\ \midrule
        \textbf{Ours} & \textbf{70.71} & \textbf{20.522} & \textbf{90.50} & \textbf{50.92} \\
    \bottomrule
    \end{tabular}
    }
\end{table}

\begin{table}[!t]
    \centering
    \captionsetup{justification=centering}
    \caption{COMPARISONS WITH STATE-OF-THE-ART MODELS ON THE KITS (2023) DATASET.}
    \label{tab:KITS(2023)}
    \resizebox{\columnwidth}{!}{%
    \begin{tabular}{cccccc}
    \toprule
        \multirow{2}{*}{Method} & DSC\textcolor{red}{$\uparrow$} & HD95\textcolor{red}{$\downarrow$} & Kidney & Lesion & Cyst \\ 
        ~ & (\%, mean) & (mm, mean) & \multicolumn{3}{c}{DSC\textcolor{red}{$\uparrow$}(\%)} \\ \midrule
        UNet & 51.79 & 30.987 & 90.14 & 33.02 & 32.20 \\
        TransUNet & 58.31 & 29.390 & 90.94 & 42.31 & 41.68 \\
        Swin-Unet & 49.41 & 31.283 & 87.02 & 35.52 & 25.69 \\
        UCTransNet & 47.19 & 44.911 & 87.92 & 18.86 & 34.79 \\
        TransCASCADE & 59.38 & 26.173 & 91.21 & 44.76 & 42.17 \\
        Cascaded MERIT & 61.01 & 23.033 & 91.29 & 45.05 & 46.69 \\
        FSCA-Net & 46.55 & 28.441 & 90.67 & 21.83 & 27.16 \\ \midrule
        \textbf{Ours} & \textbf{64.07} & \textbf{22.770} & \textbf{91.74} & \textbf{45.18} & \textbf{55.31} \\
    \bottomrule
    \end{tabular}
    }
\end{table}

\begin{table}[!t]
    \centering
    \captionsetup{justification=centering}
    \caption{COMPARISONS WITH STATE-OF-THE-ART MODELS ON THE LUNG DATASET.}
    \label{tab:LUNG}
    \resizebox{\columnwidth}{!}{%
    \begin{tabular}{ccccc}
    \toprule
        \multirow{2}*{Method}& DSC$\textcolor{red}{\uparrow}$& HD95$\textcolor{red}{\downarrow}$& Recall$\textcolor{red}{\uparrow}$& Precision$\textcolor{red}{\uparrow}$\\
        ~& (\%, mean)& (mm, mean)& (\%, mean)& (\%, mean)\\ \midrule
        U-Net& 70.50& 37.913&	66.07&	80.04\\
        TransUNet& 70.72&	25.218&	60.56&	82.06\\
        SwinUnet& 63.41&	25.131&	57.47&	77.29\\
        UCTransNet& 70.75&	32.384&	65.90&	80.38\\
        TransCASCADE & 68.31 & 28.551 & 65.01 & 76.51\\
        Cascaded MERIT & 70.43 & 28.554 & 67.33 & 83.34\\
        FSCA-Net & 72.10 & 25.545 & 68.07 & 82.23\\ \midrule
        \textbf{Ours} & \textbf{74.49} & \textbf{23.505} & \textbf{71.06} & \textbf{83.52} \\
    \bottomrule
    \end{tabular}
    }
\end{table}

\subsubsection{Experiments on KiTS (2023)} As shown in Table \ref{tab:KITS(2023)}, traditional networks perform poorly in tumor and cyst segmentation, with the best TransUNet achieving a DSC of only 42.31\% for tumors and 42.68\% for cysts. In contrast, our model improves the DSC for cyst segmentation to 55.31\%, and MCFNet excels in segmenting kidneys, tumors, and cysts.

As shown in the second row of Fig. \ref{fiveDatasets}, cyst segmentation is more challenging due to small target areas. Models like TransUNet, and Swin-Unet struggle with cyst segmentation. While UCTransNet correctly identifies cyst regions, it has difficulty maintaining accurate tumor segmentation simultaneously.

\begin{table}[!t]
    \centering
    \captionsetup{justification=centering}
    \caption{COMPARISONS WITH STATE-OF-THE-ART MODELS ON THE CPCGEA DATASET.}
    \label{tab:CPCGEA}
    \begin{tabular}{ccccc}
    \toprule
        \multirow{2}*{Method}& DSC$\textcolor{red}{\uparrow}$& HD95$\textcolor{red}{\downarrow}$& Recall$\textcolor{red}{\uparrow}$& Precision$\textcolor{red}{\uparrow}$\\ 
        ~& (\%, mean)& (mm, mean)& (\%, mean)& (\%, mean)\\ \midrule
        U-Net& 64.05& 9.312&	70.34&	65.99\\
        TransUNet& 63.18&	8.517&	70.02&	66.14\\
        SwinUnet& 60.94&	7.185&	66.20&	65.24\\
        UCTransNet& 60.25&	11.374&	67.59&	64.78\\
        TransCASCADE & 65.44 & 6.646 & 68.85 & 69.14\\
        Cascaded MERIT & 65.50 & 7.738 & 70.47 & 68.57\\
        FSCA-Net & 65.24 & 6.792 & 68.88 & 68.37\\ \midrule
        \textbf{Ours} & \textbf{67.28} & \textbf{6.064} & \textbf{71.48} & \textbf{69.43} \\
    \bottomrule
    \end{tabular}
\end{table}

\begin{table}[!t]
    \centering
    \captionsetup{justification=centering}
    \caption{COMPARISONS WITH STATE-OF-THE-ART MODELS ON THE KNEE DATASET.}
    \label{tab:KNEE}
    \begin{tabular}{ccccc}
    \toprule
        \multirow{2}*{Method}& DSC$\textcolor{red}{\uparrow}$& HD95$\textcolor{red}{\downarrow}$& Recall$\textcolor{red}{\uparrow}$& Precision$\textcolor{red}{\uparrow}$\\ 
        ~& (\%, mean)& (mm, mean)& (\%, mean)& (\%, mean)\\ \midrule
        U-Net& 98.99& 1.089&	98.65&	99.23\\
        TransUNet& 98.92&	1.094&	98.74&	99.12\\
        SwinUnet& 98.25&	1.228&	97.79&	98.72\\
        UCTransNet& 99.08&	1.055&	98.82&	99.01\\
        TransCASCADE & 99.02 & 1.000 & 98.89 & 99.16\\
        Cascaded MERIT & 99.09 & 1.000 & 98.95 & 99.24\\
        FSCA-Net & 97.59 & 5.362 & 97.00 & 98.43\\ \midrule
        \textbf{Ours} & \textbf{99.14} & 1.000 & \textbf{98.99} & \textbf{99.29} \\
    \bottomrule
    \end{tabular}
\end{table}

\subsubsection{Experiments on LUNG} We evaluate segmentation performance using four metrics, as summarized in Table \ref{tab:LUNG}. Among the baselines, FSCA-Net achieves the best DSC and Recall, while Cascaded MERIT excels in Precision at 82.23\%. Our model outperforms all models in DSC, HD95, Recall, and Precision. Specifically, MCFNet improves DSC by 2.39\% and Recall by 2.99\% compared to FSCA-Net.

As shown in the third row of Fig. \ref{fiveDatasets}, most models correctly identify tumor regions, except Swin-Unet. UNet includes background areas as tumors, while Swin-Unet misidentifies both tumor and vascular regions as tumors.

\begin{figure}[!t]
\centerline{\includegraphics[width=1.0\linewidth]{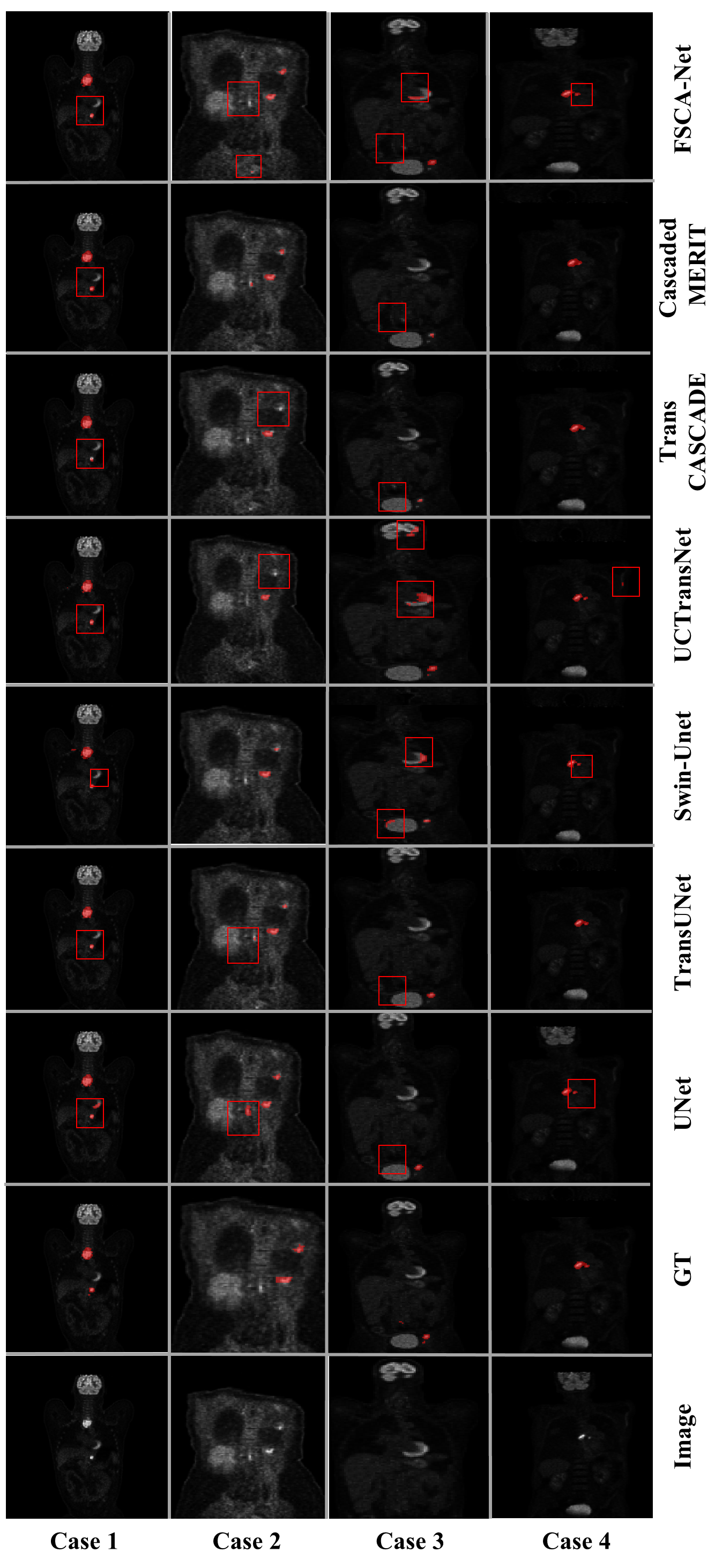}}
\caption{Visual examples of segmentation on the AutoPET dataset. The tumor lesion region is marked in red for clarity.}
\label{autopetComparedModel}
\end{figure}

\begin{figure}[!t]
\centerline{\includegraphics[width=0.8\linewidth]{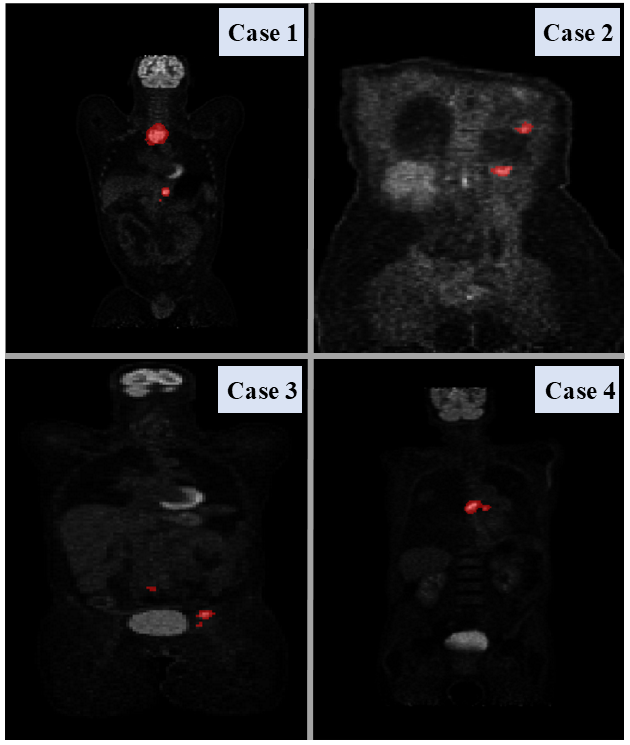}}
\caption{The visual results of our method on the AutoPET dataset. The tumor lesion region is marked in red for clarity.}
\label{autopetMCFNet}
\end{figure}

\begin{table}[!t]
    \centering
    \captionsetup{justification=centering}
    \caption{COMPARISONS WITH STATE-OF-THE-ART MODELS ON THE AUTOPET DATASET.}
    \label{tab:AUTOPET}
    \begin{tabular}{ccccc}
    \toprule
        \multirow{2}*{Method}& DSC$\textcolor{red}{\uparrow}$& HD95$\textcolor{red}{\downarrow}$& Recall$\textcolor{red}{\uparrow}$& Precision$\textcolor{red}{\uparrow}$\\ 
        ~& (\%, mean)& (mm, mean)& (\%, mean)& (\%, mean)\\ \midrule
        U-Net& 58.49& 30.832&	59.82&	68.62\\
        TransUNet& 61.95&	22.329&	60.22&	72.98\\
        SwinUnet& 50.55&	31.925&	53.67&	57.58\\
        UCTransNet& 58.05&	28.279&	59.77&	65.51\\
        TransCASCADE & 55.18 & 24.160 & 49.92 & 73.31\\
        Cascaded MERIT & 58.32 & 23.407 & 57.10 & 69.09\\
        FSCA-Net & 61.53 & 26.707 & 62.07 & 70.76\\ \midrule
        \textbf{Ours} & \textbf{63.87} & \textbf{22.251} & \textbf{65.35} & \textbf{73.91} \\
    \bottomrule
    \end{tabular}
\end{table}

\subsubsection{Experiments on CPCGEA} As shown in Table \ref{tab:CPCGEA}, we evaluate seven models for prostate cancer segmentation using four metrics. The proposed method attains the best overall performance, exceeding the previous state of the art by 1.78\% in DSC, reducing HD95 by 0.582 mm, and yielding gains of 1.01\% in Recall and 0.29\% in Precision.

As shown in the fourth row of Fig. \ref{fiveDatasets}, the comparative models fail to clearly delineate the boundaries of the tumor region. TransCASCADE and Cascaded MERIT are prone to mistakenly identifying background areas as lesions.

\subsubsection{Experiments on KNEE} As shown in Table \ref{tab:KNEE}, all models achieve an average DSC for femur segmentation exceeding 97.5\%, with MCFNet achieving the highest at 99.14\%. Furthermore, HD95, Recall, and Precision also achieve optimal performance, although the improvement in segmentation performance compared to advanced models is relatively limited.

As depicted in the fifth row of Fig. \ref{fiveDatasets}, we can see that all models accurately segment the femur region. However, Swin-Unet and UCTransNet exhibit minor segmentation errors in very small areas, although the overall differences are minimal.

\subsubsection{Experiments on AutoPET} For whole-body tumor segmentation, our designed model and seven comparison models are evaluated using four metrics, as shown in the Table \ref{tab:AUTOPET}. TransUNet achieves the best DSC of 61.95\%, the best HD95 of 22.329mm, and the best Precision of 72.98\%. Additionally, FSCA-Net achieves a Recall of 62.07\%. However, our designed model MCFNet improves by 1.92\% in DSC, reduces HD95 by 0.078mm, increases Recall by 3.28\%, and improves Precision by 0.93\%.

Fig. \ref{autopetComparedModel}  and Fig. \ref{autopetMCFNet} present four segmentation cases from the AutoPET dataset, with erroneous areas marked in red boxes. In the whole-body tumor segmentation images, there are many background areas that appear white and have smooth edges but are not tumors, leading the segmentation models to mistakenly identify tumor regions.  Most models confuse background with tumor regions, with UNet and TransUNet showing over- and under-segmentation. Swin-Unet, UCTransNet, and FSCA-Net also exhibit errors, while TransCASCADE and Cascaded MERIT show under-segmentation. MCFNet performs well but can still improve in segmenting detailed tumor boundaries.

\begin{table*}[!t]
    \centering
    \captionsetup{justification=centering}
    \caption{ABLATION EXPERIMENTS ON SEGTHOR, CHAOS, LITS, KITS2019, AND CPCGEA DATASETS.}
    \label{tab:ABLATION}
    \resizebox{\linewidth}{!}{
    \begin{tabular}{ccccccccccccc}
    \toprule
        \multirow{3}{*}{ Method } & \multicolumn{2}{c}{ SegTHOR } & \multicolumn{2}{c}{ CHAOS } & \multicolumn{2}{c}{ LiTS } & \multicolumn{2}{c}{ KiTS19 } & \multicolumn{4}{c}{ CPCGEA } \\ 
        ~& DSC$\textcolor{red}{\uparrow}$& HD95$\textcolor{red}{\downarrow}$& DSC$\textcolor{red}{\uparrow}$& HD95$\textcolor{red}{\downarrow}$& DSC$\textcolor{red}{\uparrow}$& HD95$\textcolor{red}{\downarrow}$& DSC$\textcolor{red}{\uparrow}$& HD95$\textcolor{red}{\downarrow}$& DSC$\textcolor{red}{\uparrow}$& HD95$\textcolor{red}{\downarrow}$& Recall$\textcolor{red}{\uparrow}$& Precision$\textcolor{red}{\uparrow}$\\
        ~& (\%, mean)& (mm, mean)& (\%, mean)& (mm, mean)& (\%, mean)& (mm, mean)& (\%, mean)& (mm, mean)& (\%, mean)& (mm, mean)& (\%, mean)& (\%, mean)\\ \midrule
        SEB & 78.74 & 5.063 & 90.17 & 6.051 & 66.47 & 26.199 & 67.32 & 33.096 & 63.18 & 8.520 & 70.02 & 66.14 \\
        FCB & 80.75 & 4.947 & 89.83 & 9.277 & 68.54 & 23.897 & 60.59 & 41.617 & 65.24 & 6.792 & 68.88 & 68.37 \\
        SEB+Adaptive-MFA & 79.72 & 5.220 & 90.44 & 4.366 & 67.75 & 35.554 & 67.84 & 30.069 & 63.75 & 7.746 & 75.76 & 59.78 \\
        SCB+Adaptive-MFA & 81.43& 5.040 & 90.01 & 8.253 & 70.56 & 25.951 & 64.31 & 43.377 & 65.44 & 6.646 & 68.85 & 69.14 \\
        SEB+FCB  & 82.87 & 5.306 & 91.91 & 6.001 & 73.25 & 26.803 & 68.32 & 23.997 & 65.50 & 7.738 & 70.47 & 68.57 \\\midrule
        \textbf{SEB+FCB+Adaptive-MFA} & \textbf{83.27} & \textbf{4.187} & \textbf{92.80} & \textbf{3.676} & \textbf{73.79} & \textbf{23.326} & \textbf{70.71} & \textbf{20.522} & \textbf{67.28} & \textbf{6.064} & \textbf{71.48} & \textbf{69.43} \\
    \bottomrule
    \end{tabular}
    }
\end{table*}

\begin{table}[!t]
    \centering
    \captionsetup{justification=centering}
    \caption{COMPLEXITY COMPARISON BETWEEN MLP AND LINEAR LAYERS IN THE LAT MODULE.}
    \label{tab:LAT}
    \begin{tabular}{ccccc}
    \toprule
        \multirow{2}*{Method}& FLOPs$\textcolor{red}{\downarrow}$& Params$\textcolor{red}{\downarrow}$\\ 
        ~& (G)& (M)\\ \midrule
        LAT w/ MLP &  61.2 & 57.7\\ \midrule
        \textbf{LAT w/ Linear} &  \textbf{59.3} & \textbf{47.0} \\
    \bottomrule
    \end{tabular}
\end{table}

\subsection{Ablation Study}
To validate the effectiveness of our proposed module, we conduct experiments on five datasets and use multiple evaluation metrics to comprehensively assess the module's effectiveness. As shown in Table \ref{tab:ABLATION}, after applying the adaptive loss aggregation strategy to SEB and FCB, the segmentation performance significantly improves across all datasets. Additionally, compared to using SEB and FCB individually, cascading SEB and FCB enhances the model's segmentation performance, and further applying the adaptive loss aggregation strategy on top of this results in the best segmentation performance.

\subsection{Discussion}

\subsubsection{Discussion on the Complexity Optimization of the LAT Module}
As shown in Table \ref{tab:LAT}, replacing the MLP layer with a linear layer in the LAT module effectively reduces model complexity. Specifically, the number of parameters decreases from 57.7M to 47.0M, and the computational cost (FLOPs) is reduced from 61.2G to 59.3G. The structural simplification is achieved by removing redundant non-linear transformations present in the MLP. Consequently, this optimization yields a more efficient and practical LAT module, making it particularly suitable for resource-constrained application scenarios.

\begin{table}[!t]
    \centering
    \captionsetup{justification=centering}
    \caption{ACCURACY (DICE SCORE) VS. MODEL COMPLEXITY (PARAMETERS AND COMPUTATIONAL COMPLEXITY) COMPARISON ON THE CPCGEA DATASET.}
    \label{tab:MODEL COMPLEXITY}
    \begin{tabular}{ccccc}
    \toprule
        \multirow{2}*{Method}& DSC$\textcolor{red}{\uparrow}$& FLOPs$\textcolor{red}{\downarrow}$& Params$\textcolor{red}{\downarrow}$\\ 
        ~& (\%, mean)& (G)& (M)\\ \midrule
        U-Net& 64.1& 50.3&	14.8\\
        TransUNet& 63.2& 56.7&	105.0\\
        SwinUnet& 60.9&	67.3&	82.3\\
        UCTransNet& 60.3&	63.2&	65.6\\
        TransCASCADE & 65.4& 22.1 & 123.5\\
        Cascaded MERIT & 65.5& 33.3 & 147.9\\
        FSCA-Net & 65.2 & 32.7 & 44.2\\ \midrule
        \textbf{Ours} & \textbf{67.3} & \textbf{59.3} & \textbf{47.0} \\
    \bottomrule
    \end{tabular}
\end{table}

\subsubsection{Discussion on the Model Complexity}
MCFNet integrates low parameter count, low computational cost, and high segmentation performance. Our designed model effectively balances performance with parameter and computational cost, demonstrating significant advantages over existing state-of-the-art methods. To substantiate the approach, we evaluate on the CPCGEA dataset, comparing our model with several strong baselines. Figure~\ref{modelParamFLOPs} and Table~\ref{tab:MODEL COMPLEXITY} clearly indicate that MCFNet achieves excellent segmentation performance alongside lower parameter count and computation than competing approaches. These results indicate that our model delivers higher computational efficiency and superior scalability in real-world deployments. The optimization of performance, computational cost, and parameter size makes our model suitable for resource-constrained environments and lays a solid foundation for its future application in various scenarios.

\section{Conclusion}
\label{sec:conclusion}
We design a novel multi-scale cascaded fusion network called MCFNet. This network cascades two backbones, FCB and SEB, to capture rich features from multi-scale and multi-resolution images. Among them, we introduce an innovative Linear Attention Transformer module that ensures efficient feature extraction while reducing computational and storage requirements. Finally, we propose an adaptive loss aggregation strategy called Adaptive-MFA to optimize model training and performance. Extensive experiments on multiple benchmarks demonstrate that MCFNet achieves high segmentation performance with strong generalization.

MCFNet delivers efficient, high-accuracy segmentation, improving medical image analysis and enhancing clinical precision. By accurately segmenting OARs and tumor regions, it assists doctors in formulating better treatment plans, thereby improving patient outcomes. Its strong generalization ability allows MCFNet to adapt to data generated by different hospitals and equipment, making it a promising tool for widespread clinical application and advancing medical image analysis.

% \bibliography{sn-bibliography}

\vfill

\end{document}